\documentclass[sigconf]{acmart}

\renewcommand\footnotetextcopyrightpermission[1]{} 
\makeatletter
\renewcommand\@formatdoi[1]{\ignorespaces}
\def\mdseries@tt{m}
\makeatother
\usepackage[plain]{fancyref}
\usepackage[draft=true]{minted}
\usepackage{color}
\usepackage{hyperref}
\hypersetup{
    colorlinks=true,
    linkcolor=blue,
    filecolor=red,      
    urlcolor=magenta,
    breaklinks=true,
}
\usepackage{breakurl}

\usepackage{amssymb}
\usepackage[ruled,vlined]{algorithm2e}
\usepackage{enumitem}
\usepackage{graphicx}
\usepackage{subcaption}
\usepackage[listings,skins]{tcolorbox}
\usepackage{balance}

\setcopyright{none}

\newcommand{\ovaled}[1]{%
  \tikz[baseline=(char.base)]\node[anchor=south west, draw,rectangle, rounded corners, inner sep=1.2pt, 
    ](char){
    #1} ;}

\def\PMemory{Persistent Memory~}
\def\PMemoryN{Persistent Memory}
\def\PM{PM~}
\def\PMN{PM}
\def\PMCN{Persistent Memory Controller}
\def\PMC{\PMCN~}
\def\indent{\hspace{5pt}}
\def\indentA{\hspace{20pt}}
\def\indentB{\hspace{40pt}}

\begin{document}
\sloppy

\title{Hardware Transactional Persistent Memory}

\author{Ellis Giles}
\affiliation{
  \institution{Rice University}
  \city{Houston}
  \state{TX}
  \postcode{77005} 
  \country{USA} }
\email{erg@rice.edu}
\author{Kshitij Doshi}
\affiliation{%
  \institution{Intel Corporation}
  \city{Chandler}
  \state{AZ}
  \postcode{85226} 
  \country{USA} }
\email{kshitij.a.doshi@intel.com}
\author{Peter Varman}
\affiliation{%
  \institution{Rice University}
  \city{Houston}
  \state{TX}
  \postcode{77005} 
  \country{USA} }
\email{pjv@rice.edu}

\begin{abstract}
Emerging Persistent Memory technologies (also {\small PM}, Non-Volatile DIMMs, Storage Class Memory or {\small SCM}) 
hold tremendous promise for accelerating popular 
data-management applications like in-memory databases. 
However, programmers now need to deal with ensuring the atomicity of transactions
on Persistent Memory resident 
data and maintaining consistency between 
the order in which processors perform stores and that in which the updated values become durable.

The problem is specially challenging when high-performance isolation mechanisms like 
Hardware Transactional Memory ({\small HTM}) are used for concurrency control.
This work shows how {\small HTM} transactions can be ordered correctly and 
atomically into \PM by the use of a novel software protocol combined 
with a \PMCN, without requiring changes to processor 
cache hardware or {\small HTM} protocols. 
In contrast, previous approaches require significant changes to existing processor 
microarchitectures.
Our approach, evaluated using both micro-benchmarks and the {\small STAMP} suite
compares well with standard (volatile) {\small HTM} transactions.
It also yields significant gains in throughput and latency in comparison with 
persistent transactional locking.

\end{abstract}

\maketitle

\section{Introduction}

This paper provides a solution to the problems of adding {\it durability} to concurrent
in-memory transactions that use Hardware Transactional Memory ({\small HTM}) for
concurrency control while operating on data in byte-addressable Non-Volatile Memory or Persistent Memory ({\small PM}).

Recent years have witnessed a sharp shift towards real time data-driven
and high-throughput applications.
This has spurred a broad adoption of in-memory and massively parallelized
data processing~\cite{saphana, blu13vldb, scispark} across business,
scientific, and industrial application domains~\cite{gridgain, sparkML}.
Two emerging hardware developments provide further impetus to this approach
and raise the potential for transformative gains in speed and scalability:
(a) the arrival of inexpensive, byte-addressable, and large capacity
Persistent Memory devices
~\cite{intel3dxp}
eliminates the I/O operations and bottlenecks
common to many data management applications, and,
(b) the availability of CPU-based transaction support (with
{\small HTM})~\cite{htm, sle}) makes it straight-forward for threads to work
spontaneously in shared memory spaces without having to synchronize explicitly.

To prevent corruption of state upon an untimely machine or software failure,
a sequence of store operations to \PM in a transactional section of code
cannot be partially reflected into \PMN;
nor can it be transmitted piecemeal from processor caches into \PM without
similarly risking significant loss of data.
Operating on \PMemory based data thus produces new atomicity and consistency requirements.
Software approaches for ensuring atomic durable updates share some
characteristics with {\small HTM} techniques in commercial processors --
both checkpoint state at some level of granularity and guard against
communication of partial updates.
However, different mechanisms are at play: while stable stores to persistent
media are usually obtained by covering updates with 
logging or versioning,
partial updates are prevented from being propagated between
threads in {\small HTM} transactions by {\small CPU}s sheltering them
until the transaction closes.
Once an {\small HTM} transaction closes, its updates become visible en masse
through the cache hierarchy and can travel in any order to memory {\small DIMM}s.
However stable storage of the updates into \PMemory by the transaction
additionally requires an ability to reliably delineate its updates from those
by other overlapping transactions, and to use that delineation to recover
from an unanticipated machine restart.

Existing \PM programming frameworks separate into categories based on the
degree of change they require in the processor architecture for such problems.
Several works
~\cite{volos11, msst15, atlas14, rewind, dudetm}
operate with existing processor capabilities and
write a log (either a write-ahead log or an undo log) durably to cover data
changes arriving via the volatile cache hierarchy;
some
~\cite{zhao13, atom, spms}
require significant changes to existing cache hardware and protocols in the processor
microarchitecture, while 
others ~\cite{cf13, cal16, doshi16} only require external controllers 
not affecting the processor core.

Isolation among concurrent transactions in the above works is achieved
either by the use of two-phase locking protocols or provided within the rubric of an
{\small STM} ~\cite{volos11}.
Logging based software approaches are problematic for {\small HTM}
transactions (e.g., Intel {\small TSX}) which cannot bypass the caches
in order to flush the log records synchronously into \PMemory ahead of transaction closings.
For these, {\small PTM}~\cite{ptm}
proposes changes to processor caches while adding an on-chip
scoreboard and global transaction id register to couple HTM with \PMN.
Recent work~\cite{avni1,avni2,dudetm} has attempted to provide
inter-transactional isolation by employing processor
based ({\small HTM})~\cite{htm, sle} mechanisms.
However, these solutions all require changes to the existing {\small HTM} semantics
and implementations--
~\cite{avni1} and ~\cite{avni2} propose a new
instruction to perform non-aborting cacheline flush from
within an {\small HTM}, while ~\cite{dudetm}
proposes allowing non-aborting concurrent access to designated memory variables
within an {\small HTM}.

Selective and incremental changes to the clean isolation semantics of {\small HTM}
are not to be undertaken lightly; understanding their impact on global system
correctness and performance typically requires long gestation periods before
processor manufacturers will embrace them.
In this paper we provide a new solution to obtain persistence consistency in
\PMemory while using {\small HTM} for concurrency control.
The solution does not alter the processor microarchitecture, but leverages
a very simple, external \PMemory controller along with a persistence
protocol, to supplement the existing {\small HTM} semantics and allowing {\small HTM}
transactions to operate at the speed of in-memory volatile transactions.
The solutioni, while it achieves the concurrency benefits of {\small HTM} for
\PMN-based data also applies to non-{\small HTM} transactions in a
straight-forward way.

\section{Overview}

\subsection{{HTM+PM}  Basics}

Hardware Transactional Memory, or HTM, was introduced in ~\cite{htm} as a new, easy-to-use method for lock-free synchronization supported by hardware.
The initial instructions for HTM included load and store transactional instructions in addition to transactional management instructions.
Most HTM implementations extend an underlying cache-coherency protocol to handle detection of transactional conflicts during program execution.
The hardware system performs a speculative execution on a demarcated region of code similar to an atomic section. Independent transactions (those that do not write a shared variable) proceed unrestricted through their HTM sections.
Transactions which access common variables concurrently in their HTM sections, with at least one transaction performing a write, are serialized by the HTM. That is, all but one of the transactions is aborted; an aborted transaction will restart its HTM code at the beginning.
Updates made within the HTM section are hidden from other transactions and are prevented from writing to  memory until the transaction successfully completes the HTM section.
This mechanism provides atomicity (all-or-nothing) semantics for individual transactions with respect to visibility by other threads, and serialization of conflicting, dependent transactions.
However HTM was originally designed for volatile memory systems (rather than supporting database style ACID transactions) and therefore any power failure leaves main memory in an unpredictable state relative to the actual values of the transaction variables.

Persistent Memory, or {\small PM}, introduces a new method of persistence to the processor.  \PMN, in the form of persistent DIMM, resides on the main-memory bus alongside {\small DRAM}. Software can access persistent memory using the usual {\small LOAD} and {\small STORE} instructions used for {\small DRAM}. Like other memory variables, \PM
variables are subject to forced and involuntary cache-evictions and encounter other deferred  memory  operations done by  the processor. 

For Intel {\small CPU}s, {\small CLWB} and 
{\small CLFLUSHOPT}  instructions provide the ability to flush modified data (at cacheline granularity) to be evicted from the processor cache hierarchy.
These instructions, however, are weakly ordered with respect to other store operations in the instruction stream. Intel has extended the semantic for SFENCE to cover such flushed store operations so that software can issue SFENCE to prevent new stores from executing until previously flushed data has entered a power-safe domain; i.e., the data is guaranteed by hardware to reach its locations in the \PM media.
This guarantee also applies to data that is written to \PM with instructions that bypass the processor caches.
However, when executing within an HTM transaction, a CPU cannot exercise CLWB, CLFLUSHOPT, non-cacheable stores, and SFENCE instructions since the stores by the CPU are considered speculative until the HTM transaction completes successfully.

Even though HTM guarantees that transactional values are only visible on transaction completion, hardware manufacturers cannot simply utilize a non-volatile processor cache hierarchy or battery backed flushing of the cache on failures to provide transactional atomicity. Transactions that do not complete before a software or hardware restart produce partial and therefore inconsistent updates in non-volatile memory, as there is no guarantee when a machine halt will occur.  The halt may happen during XEND execution leaving only partial updates in cache or write buffers which can corrupt in-memory data structures.

\subsection{Challenges of persistent HTM transactions}

Consider  transactions $A$, $B$ and $C$ shown in Listings~\ref{listing:ta},~\ref{listing:tb} and~\ref{listing:tc}. 
Assume that  {\bf w}, {\bf x}, {\bf y}, {\bf z}  are persistent variables initialized to zero in their home locations in \PMemory (\PMN).
The code section demarcated between the instructions  {\bf {\small XB}egin} and {\bf {\small XE}nd}  will be referred to as an 
{\it {\small HTM} transaction} or simply a transaction.
The {\small HTM} mechanism ensures the {\it atomicity} of  transaction execution.
Within an  {\small HTM} transaction, all updates are made to private locations in the cache, and the hardware guarantees that the 
updates are {\it not allowed} to propagate to their home locations in \PMN.
After the {\bf {\small XE}nd} instruction completes, all of the cache lines updated in the transaction become instantaneously 
visible in the cache hierarchy. 

\smallskip
\noindent
{\bf Atomic Persistence}:
The first challenge is to ensure that the transaction's updates that were made
atomically in the cache are also persisted atomically in \PMN.
Following   {\bf {\small XE}nd},  the transaction variables are once again subject to the normal cache operations like evictions and the use of cache write-back instructions.
There are no  guarantees regarding whether or when the transaction's updates 
actually get written to \PM from the cache. This can create a problem if  the machine crashes  before all these updates are written back to \PMN.
On a reboot, the  values of these variables in \PM will be inconsistent with the pre-crash transaction values.
This leads to the first requirement: 

\noindent
\begin{itemize}
\item 
{\it Following crash recovery, ensure that all or none of the updates of an {\small HTM}  transaction are stored in their \PM home locations.}
\end{itemize}

A common solution is to log the transaction updates in a separate persistent storage area before allowing  them to update their \PM home locations.
Should a crash interrupt the updating of the home locations, the saved log can be replayed. When transactions execute within an {\small HTM} there is a problem with this solution since the log cannot be written to \PM within the transaction and can be done only after the {\bf {\small XE}nd}.
At that time the transaction updates are also made visible in the cache hierarchy and are susceptible to uncontrolled cache evictions into \PMN.
Hence there is no guarantee that the log has been persisted before transaction updates have percolated into \PMN.
We describe our solution in Section~\ref{sec:solution}.

\begin{figure}
\noindent\begin{minipage}{.14\textwidth}
\begin{lstlisting}[caption={ A},language=C,frame=tlrb,label={listing:ta}]
A() {
 XBegin;
  w = 1;
  x = w;
 XEnd;
}
\end{lstlisting}
\end{minipage}\hfill
\begin{minipage}{.14\textwidth}
\begin{lstlisting}[caption={B},language=C,frame=tlrb,label={listing:tb}]
B() {
 XBegin;
  w = w+1;
  y = w;
 XEnd;
}
\end{lstlisting}
\end{minipage}\hfill
\begin{minipage}{.14\textwidth}
\begin{lstlisting}[caption={C},language=C,frame=tlrb,label={listing:tc}]
C() {
 XBegin;
  w = w+1;
  z = w;
 XEnd;
}
\end{lstlisting}
\end{minipage}\hfill
\vspace{-10pt}
\end{figure}

\smallskip
\noindent
{\bf Persistence Ordering}: The second problem deals with ensuring that the {\it execution order} of dependent  {\small HTM} 
transactions is correctly reflected in \PM following crash recovery. As an example, consider the dependent transactions $A, B, C$ in Listings 1, 2 \& 3.
The {\small HTM} will serialize their execution in some order: say  $A$, $B$ and $C$.
The values of the transaction variables following the execution of $A$ are given by the vector $V_1$ = [{\bf w, x, y, z}] = [$1, 1, 0, 0$]; after the execution of $B$ the vector becomes $V_2$ = [$2, 1, 2, 0$] and finally following $C$ it is
$V_3$ = [$3, 1, 2, 3$].
Under normal operation the write backs of variables to \PM from different transactions may become arbitrarily interleaved.
For instance suppose that
$x$ is evicted immediately after $A$ completes, $w$ after $B$ completes, and $z$ after $C$ completes.
The persistent memory state is then
[$2, 1, 0, 3$]; should the machine crash, the \PM will contain this meaningless combination of values on reboot. 
A consistent state would be either the initial vector [$0, 0, 0, 0$] or one of $V_1, V_2$ or $V_3$.
This leads to the second requirement: 

\noindent
\begin{itemize}
\item 
{\it  {\it Following crash recovery, ensure that the persistent state of any sequence of dependent transactions  is consistent with their execution order.}}
\end{itemize}

If  individual transactions satisfy atomic persistence, then it is sufficient to  ensure that \PM is updated in transaction execution order. With software concurrency control (using an {\small STM} or two-phase transaction locking), it is straightforward to correctly order the updates simply by saving a transaction's log {\it before} it commits  and releases its locks.
In case of a crash, the saved logs are simply replayed in the order they were saved, thereby reconstructing  persistent state to a correctly-ordered prefix of the executed transactions. 

When {\small HTM} is used for concurrency control the logs can only be written to \PM {\it after} the transaction {\bf {\small XE}nd}.
At that time other dependent transactions can execute and race with the completed transaction, perhaps writing out their logs before the first.  Solutions like using an atomic counter within  transactions to order them correctly are not practical since the shared counter will result  in {\small HTM}-induced aborts and serialization of all transactions.  Some papers have advocated that processor manufacturers alter
 {\small HTM} semantics and implementation to allow selective writes to \PM from 
 within an {\small HTM}~\cite{avni1,avni2,dudetm}.
We describe our solution without the need for such intrusive processor changes in Section~\ref{sec:solution}.

\smallskip
\noindent
{\bf Strict and Relaxed Durability}:
In traditional {\small ACID} databases, a committed transaction is guaranteed to be durable since its log is made persistent before it commits. We  refer to this property as {\it strict durability}. In {\small HTM} transactions the log is written to \PM {\it after} the 
{\bf {\small XE}nd} instruction some time before the transaction commits. A natural question is to characterize the time 
it is safe for a transaction requiring strict durability to commit.

It is generally not safe to commit a transaction $Y$ at the time it completes persisting its log for the same reason that it is difficult to ensure persistence ordering.
Due to races  in the code outside the {\small HTM}  it is possible that an earlier  transaction $X$ (on which  $Y$ depends) to have completed but not yet persisted its log.
When recovering from a crash that occurs at this time, the log of $Y$ should not be replayed since earlier transaction $X$ cannot be replayed.
This leads to the third requirement: 

\noindent
\begin{itemize}
\item 
{\it  {\it Following crash recovery, strict durability requires that every committed transaction is persistent in \PMN.}}
\end{itemize}

We define a new property known as {\it relaxed durability} that allows
an individual transaction to  opt for an early commit immediately after it persists its log. Requiring relaxed or strict durability is 
a local choice made individually by a  transaction based on the application requirements. Transactions choosing relaxed durability 
face a window of vulnerability after they commit, during which a crash may result in their transaction updates 
not being reflected in \PM  after recovery. The gain is potentially reduced transaction latency.
However, irrespective of the choice of the durability model by individual transactions, the underlying persistent memory will always recover to a consistent state, reflecting an ordered atomic prefix of every sequence of dependent transactions.

\section{Our Approach}
\label{sec:solution}
Our approach achieves durability of HTM transactions to Persistent Memory by a
cooperative protocol involving three components: 
 a back-end \PMC, transaction execution and logging, and a failure recovery procedure.
The \PMC intercepts dirty cache lines evicted from the last-level processor cache ({\small LLC}) 
on their way to persistent memory.
An intercepted cache line is held in a FIFO queue within the controller until it is safe to write it out to \PMN.
All memory variables are subject to the normal cache operations and are fetched and evicted 
according to normal cache protocols.
The only change we introduce is  interposing the external controller between the {\small LLC} and memory.
Note that the controller does not require changing any of the internal processor behavior. 
The controller simply delays the evicted cache lines on their way to memory till it can guarantee safety.
It is pre-programmed with the address range of a region of persistent memory that is reserved for holding transaction logs.
Addresses in the log region pass through the controller without intervention. 

{\small HTM+PM} transactions execute independently.
Within an outer-envelope that achieves consistency of updates between the volatile cache hierarchy and the durable state in \PMN, these transactions use an unmodified {\small HTM} to serialize the computational portions of conflicting transactions.
A transaction 
(1) notifies the controller when it  opens and closes,
(2) saves start and end timestamps in \PM to enable consistent recovery after a failure,
(3) performs its {\small HTM} operation, and 
(4) persists a log of its updates in \PM before closing.
If a transaction requires strict durability it informs the controller during its closing step, and then
waits for the go-ahead from the controller before committing.
If it only needs relaxed durability it can commit immediately after its close.
The recovery routine is invoked after a system crash to restore the \PM variables to a valid state
{\it i.e.} a state  that is consistent with the actual execution order of every sequence of dependent transactions.
The recovery procedure uses the saved logs to recover the values of the updated variables, and the saved start and
end timestamps to determine which logs are eligible for replay and their replay order. 

\subsection{Transaction Lifecycle}
\label{sec:solution:lifecycle}
A transaction can be viewed as progressing through five states: {\bf {\small OPEN}},  {\bf {\small COMPUTE}}, {\bf {\small LOG}}, 
{\bf {\small CLOSE}} and {\bf {\small COMMIT}} as shown in Listing ~\ref{listing:lifecycle}.
When a transaction begins, it calls the library function {\it OpenWrapC} (see Algorithm~\ref{algo:SwWrapAlgo} in Section~\ref{sec:impl:sw}).
This function invokes the \PMC with a small unique integer ({{\it wrapId}})  that identifies the transaction. 
The controller adds {\it wrapId} to a set 
of currently open transactions (referred to as {\small COT})
that it maintains (see Algorithm~\ref{algo:HwWrapAlgo} in Section~\ref{sec:impl:hw}). 
The transaction then allocates and initializes space in \PM for its log and updates  the  {\it  {startTime}} record  
of the log.
The {\it startTime} is obtained by reading  a system wide platform timer  using the  {\it \small RDTSCP} instruction (see Section~\ref{sec:impl:ts}).
In addition to {\it startTime}, a log  includes a second  timestamp  {\it persistTime} that will be set just prior to completing the 
{\small HTM} transaction.
The  {\it writeSet}  is a sequence of ($address, value$)  pairs  in the log that will be filled with 
the locations and  values of  the updates  made by the {\small HTM} transaction.  
The log  with its recorded {\it startTime}  is then  persisted using cache line write-back instructions ({\it clwb}) and {\it sfence}.

The transaction then enters the { {\small COMPUTE}} state by executing {\bf {\small XB}egin} and entering the {\small HTM} code section.
Within the {\small HTM} section, the transaction updates  {\it writeSet} with the persistent variables that it writes.
Note that the records in {\it writeSet} will be held in cache during the {\small COMPUTE} state since it occurs within 
an {\small HTM} and  cannot be propagated to \PM until {\bf {\small XE}nd} completes. Immediately before {\bf{\small XE}nd} 
the transaction obtains a second timestamp {\it {persistTime}} that will be used to order the transactions correctly.
This timestamp is also obtained using the same {\it \small RDTSCP} instruction.

After executing {\bf {\small {XE}nd}}, a transaction next enters the {{\small LOG}} state.
It flushes its log records from cache hierarchy into \PM using cache line write-back instructions ({\small CLWB} or {\small CLFLUSHOPT}), following the last of them with an {\small SFENCE}.
This ensures that all the log records have been persisted.
In addition to $startTime$, a log includes the $persistTime$ time stamp that was set just prior to completing the transaction.
The $writeSet$ records in the log hold (address; value) pairs representing the locations and the values updated by the transaction.
After the {\small SFENCE} following the log record flushes, the transaction enters the {\small CLOSE} state.

\begin{figure}[tp]
\begin{minipage}{\linewidth}
\begin{lstlisting}[caption={Transaction Structure},language=C,frame=tlrb,label={listing:lifecycle}]
Transaction Begin
----------  State: OPEN  ----------------
 1 NotifyPMController(open);
 2 Allocate a Log structure in PM; 
 3 nowTime = ReadPlatformCounter();
 4 Log.startTime = nowTime;
 5 Log.writeSet = {};
 6 Persist Log in PM;
----------  State: COMPUTE  -------------
 XBegin
 // Transaction Body of HTM
 //   All reads and writes to transaction 
 //  variables are performed and also
 //  appended to  Log.writeSet.
 7   endTime = ReadPlatformCounter();
 XEnd
	 
----------  State: LOG  -----------------
 8 Log.persistTime = endTime;
 9 Persist  Log in PM; 
    
----------  State: CLOSE  ---------------
 10 NotifyPMController(close);
 
----------  State: COMMIT  --------------
11 If (strict durability requested)
	 Wait for notification by controller;
12 Transaction End
\end{lstlisting}
\end{minipage}
\end{figure}

In the {{\small CLOSE}} state the transaction signals the \PMC that its log has been persisted in \PMN.
The controller removes the transaction from its set of currently open
transactions  {\small COT}. It also reflects the closing in the state of evicted cache lines in its FIFO buffer as described below in Section~\ref{sec:solution:controller}.
A transaction requiring strict durability informs the controller at this time; the controller will signal the transaction in due course when it is safe to commit {\it i.e.} its updates are guaranteed to be durable.

The transaction is then complete and enters the {{\small COMMIT}} state.
If it requires strict durability it waits till it is signaled by the controller.
Otherwise, it immediately commits and leaves the system.

\begin{figure}[tp]
  \includegraphics[width=0.5\textwidth]{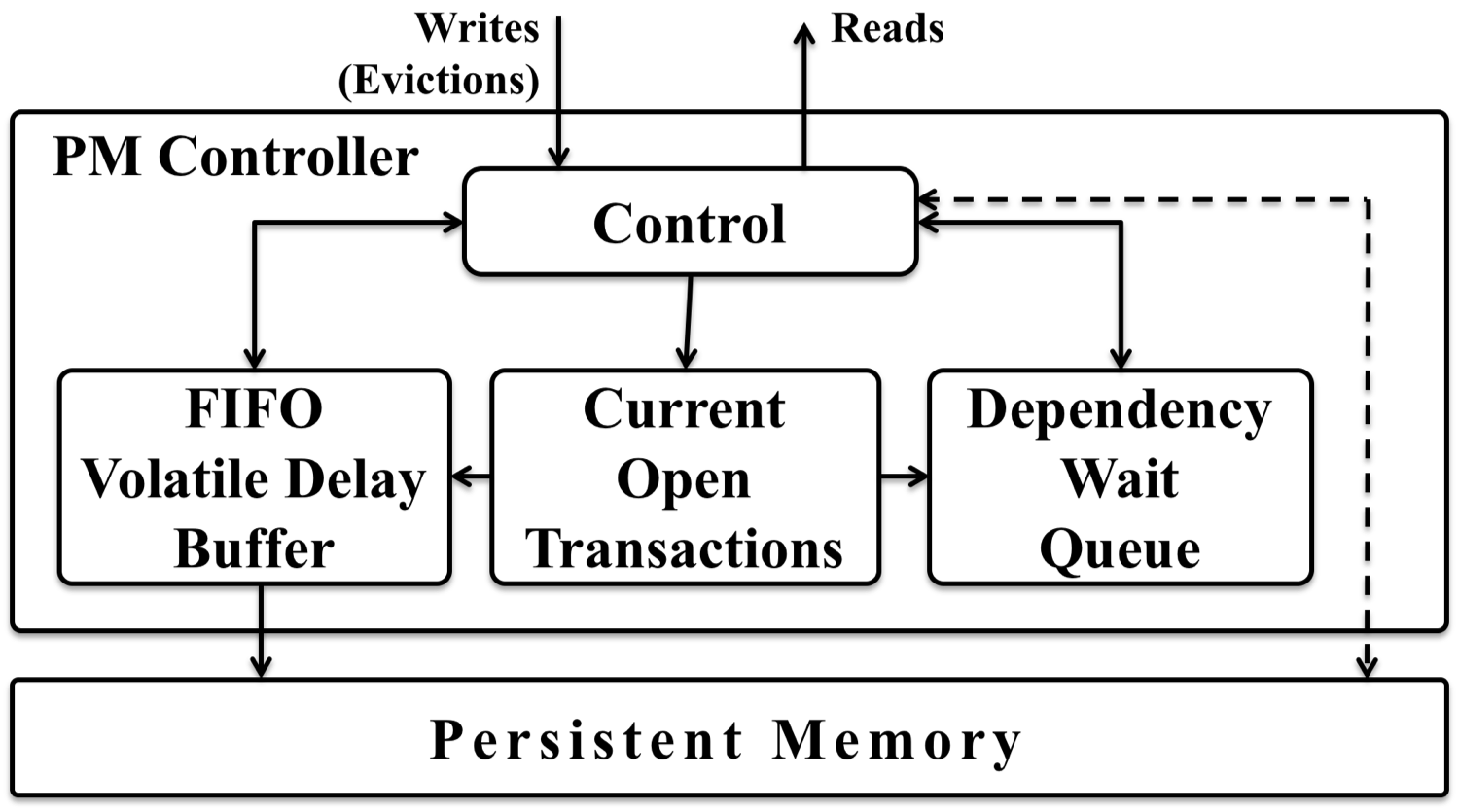}
  \caption{PM Controller with Volatile Delay Buffer, Current Open Transactions Set, and Dependency Wait Queue}
  \label{fig:controller}
\end{figure}

\subsection{\PMC}
\label{sec:solution:controller}
The \PMC is shown in Figure ~\ref{fig:controller}. 
While superficially similar to an earlier design
~\cite{spaa17, doshi16, cal16},
this controller includes enhancements to handle the subtleties of using {\small HTM} rather than
locking for concurrency control, and makes significant simplifications by shifting some of the responsibility for the maintenance of \PM state to the recovery protocol.

A crucial function of the controller is to prevent any transaction's updates from reaching \PM until it is safe for it to do so - without requiring detailed book-keeping about which transactions, currently active or previously completed, generated a specific update.
It does this by enforcing two requirements before allowing an evicted dirty cache line (representing a transaction's update) from proceeding to \PMN: (i) ensuring that the log of the transaction has been made persistent, and (ii) guaranteeing that  the saved log will be replayed during recovery. 
The first condition is needed (but not sufficient) for atomic persistence by guarding against a failure that occurs after only 
a subset of a transaction's  updates have been persisted in \PMN.
The second requirement arises because transaction logs are persisted outside the {\small HTM}, and there is no relation between the order in which the transactions execute their {\small HTM} and the order in which their logs are persisted. 
To maintain correct persistence ordering, the recovery routine may not be able to replay a log.
We illustrate the issue in the example below.

\smallskip
\noindent
{\bf Example}: Consider two dependent transactions {\small A} and {\small B},
{\bf {\small A}}:  \{w = 3; x=1;\}  and {\bf {\small B}}: \{y = w+1; z = 1;\}.
Assume that {\small HTM} transaction {\small A} executes before {\small B}, but that 
{\small B} persists its logs and closes before {\small A}.
Suppose there is a crash just after {\small B} closes. The recovery routine will not 
replay either {\small A}  or {\small B}, since the log of the earlier transaction {\small A}  is
not available. This behavior is correct.

\noindent
Now consider the situation where $y$  (value $4$) is evicted from the cache
and then written to  \PM after {\small B} persists its log.
Once again, following a crash at this time, neither log will be replayed. However, 
atomic persistence is now violated since {\small B}'s updates are partially reflected in \PMN.
Note that this violation occurred even though the write back of  $y$ to \PM 
happened after {\small B}'s log was persisted. 
The \PMC protocol prevents a write back  to \PM unless it can also
guarantee that the log of the transaction creating the update will be played back on recovery
(see Lemmas C2 and C4 in Section~\ref{sec:solution:lemmas}).

\smallskip
The second function of the controller is to track when it is safe for a transaction requiring strict durability to commit.
It is not sufficient to commit when a  transaction's logs are persistent on \PM since, as seen in the example above, the recovery routine
may not replay the log if that would violate persistency ordering.
The controller protocol effectively delays a strict durability transaction $\tau$~from committing until the earliest open transaction has a {\it startTime} greater than the {\it persistTime} of $\tau$.
This is because the recovery protocol will replay a log (see Section~\ref{sec:solution:recovery}) if and only if all transactions with {\it startTime} less than its {\it persistTime} have closed.

\smallskip
\noindent
{\bf  Implementation Overview}:

The controller tracks transactions  by maintaining a {\bf {\small COT}} (currently open transactions) set $S$.
When a transaction opens, its identifier is added to {\small  COT} and when the transaction closes it is removed.
The write into \PM of a cache line $C$ evicted into the \PMC is deferred by placing it at the tail of a {\small FIFO}  
queue maintained by the controller.
The cache line is also assigned a tag called its {\it dependency set},
 initialized with $S$ the  value of {\small COT}, at the instant that $C$ entered the \PMCN.

The controller holds the evicted instance of $C$ in a {\small FIFO} until all transactions that are in its dependency set (i.e. $S$) have closed.
When a transaction closes it is removed from both the {\small COT} and
from the dependency sets of all the {\small FIFO} entries.
When the dependency set of a cache line in the {\small FIFO} becomes empty, it is eligible to be flushed to \PMN.
One can see that the dependency sets will become empty in the order in which the cache lines were evicted, since a transaction still in the dependency set of $C$ when a
new eviction enters the {\small FIFO} will also be in the dependency set of the new entry.
The simple protocol guarantees that all transactions that opened before cache line $C$ was evicted into the controller 
(which must also include the transaction that last wrote $C$) must have closed and  persisted their logs when $C$ becomes eligible to be written to \PMN.
 This also implies that all transactions with {\it startTime} less than the
{\it persistTime}  of the transaction that last wrote {\small C} would have closed, satisfying the condition for log replay.
Hence the cache line can be safely written to \PM without violating
atomic persistence. 

Note that the evicted cache lines intercepted by the controller do not hold
any identifying transaction information and can occur at arbitrary times after the transaction leaves the 
{\small COMPUTE} state. The cache line could hold the update of a currently open transaction or
could be  from a transaction that has completed or even committed and left the system.
To guarantee safety, the controller  must perforce assume that the first situation holds. 
The details of the controller implementation  will be presented in Section~\ref{sec:impl:hw}.

\subsection{Recovery}
\label{sec:solution:recovery}

Each  completed transaction saves its log in \PMN.
The log holds  records  {\it startTime}  and {\it persistTime}   obtained by reading  the platform timer using {\small RDTSCP} instruction.
We refer to these as the start and end timestamps of  the transaction.
The start timestamp is persisted {\it before} a transaction enters its {\small HTM}. 
This allows the recovery routine to identify a transaction that started but had not completed at the
time of a failure. Note even though such a transaction has not completed, it
could still have finished  its {\small HTM} and fed values to a later dependent transaction which has completed and persisted its log.
The end timestamp and  the write set of the transaction are persisted 
{\it after} the transaction completes its {\small HTM} section, followed by an end of log marker.
There can be an arbitrary delay between the end of the {\small HTM} and the time that its log is flushed from caches into \PM and persisted.

The recovery procedure  is invoked on reboot following a  machine failure.
The routine will restore \PM values  to a consistent state that satisfies
persistence ordering by copying values from the {\it writeSet}s of the logs of  qualifying transactions to the specified addresses.
A  transaction $\tau$~ qualifies for log replay if and only if 
all earlier transactions on which it depends (both directly and transitively) are also replayed.

\smallskip
\noindent
{\bf  Implementation Overview}:

The recovery procedure first identifies the set of incomplete transactions ${\mathcal{I}}$, which have started (as indicated by the presence of a {\it startTime} record in their  log) but have not completed (indicated by the lack of a valid end-of-record marker).
The remaining complete transactions (set ${\mathcal {C}}$) are potential candidates for replay.
Denote the smallest start timestamp of transactions in ${\mathcal{I}}$  by $T_{min}$.
A transaction in ${\mathcal {C}}$ is valid (qualifies for replay) if its end timestamp ({\it persistTime}) is no more than $T_{min}$.
All valid transactions are replayed in increasing  {\it order of} their end timestamps {\it persistTime}.

\subsection{Protocol Properties} \label{sec:solution:lemmas}

We now summarize the invariants maintained by our protocol.

\medskip
\noindent
{\bf Definition}: The precedence set of a transaction $T$, denoted by {\bf prec}($T$), is the
set of all dependent transactions that executed their {\small HTM} before $T$. Since the {\small HTM} properly orders any two dependent transactions the set is well defined.

\medskip
\noindent
{\bf Lemma C1}:
Consider a  transaction $X$ with a precedence set  {\bf prec}($X$). 
For all transactions $Y$ in {\bf prec}($X$), {\it startTime}($Y$) $<$  {\it persistTime}($X$).

\smallskip
\noindent
{\bf Proof Sketch}: Let $Y$ be a transaction in {\bf prec}(X).
First let us consider direct precedence, in which a cacheline $C$ modified in $Y$ controls the ordering of $X$ with respect to $Y$.
That is, $X$ either reads or writes the cacheline $C$.
Since $Y$ is in {\bf prec}($X$), the earliest time that $X$ accesses $C$ must be no earlier than the latest time that $Y$ accesses $C$, and thus {\it persistTime}($Y$) $<$ {\it persistTime}($X$).
Next consider a chain of direct precedences, $Y$ $\rightarrow$ $Z$ $\rightarrow$ $W$ $\rightarrow \cdots X$,
which puts $Y$ in {\bf prec}($X$); and by transitivity, {\it persistTime}($Y$) $<$  {\it persistTime}($X$).  Since  {\it startTime}($Y$) $<$ {\it persistTime}($Y$)  the lemma follows.

\medskip
\noindent
{\bf Lemma C2}:
Consider  transactions $X$ and $Y$ with  {\it startTime}($Y$) $<$  {\it persistTime}($X$). 
If a cache line $C$ that is updated by $X$ is written to \PM by the controller at time $t$, then
$Y$ must have closed and persisted its log before $t$.

\smallskip
\noindent
{\bf Proof Sketch}:
Suppose $C$ was evicted to the controller at time $t' \leq t$.
Now $t'$ must be later than the time $X$ completed 
{\small HTM} execution and set  {\it persistTime}($X$);  by assumption this is after $Y$
set its {\it startTime} at which time $Y$  must have been registered as an open transaction by
the controller. Now, either $Y$ has closed before $t'$ or is still open at that time.
In the latter case, $Y$ will be added to the dependence set of $C$ at  $t'$.
Since $C$ can only be written to \PM after its  dependence set is empty,
it follows that $Y$ must have closed and removed itself from the dependence set of $C$.

\medskip
\noindent
{\bf Lemma C3}
Any  transaction $X$ that  writes an update to \PM and closes at time $t$
will be replayed by the recovery routine if there is a crash any time after $t$.

\smallskip
\noindent
{\bf Proof} 
The recovery routine will replay a transaction $X$  if the only incomplete transactions 
(started but not closed) at the time of the crash  started after $X$ completed;
that is, there is no incomplete transaction $Y$ that has a {\it startTime}($Y$)  $\leq$ {\it persistTime}($X$).
By Lemma C2  such an incomplete transaction cannot exist.

\medskip
\noindent
{\bf Lemma C4}: 

Consider a  transaction $X$ with a precedence set {\bf prec}($X$). 
Then by the time $X$ closes and persists its logs, one of the following must hold:
(i) Some update of $X$ has been written back to \PM and
all transactions $Y$ in {\bf prec}($X$)  have persisted their logs; (ii) No update of $X$ has been written to 
\PM and all transactions $Y$ in {\bf prec}($X$)  have persisted their logs; (iii) No update of $X$ has been written to 
\PM and some transactions $Y$ in {\bf prec}($X$)  have not yet persisted their logs.

\smallskip
\noindent
{\bf Proof Sketch}:
From Lemmas C1 and C2, it is not possible to have a transaction in {\bf prec}($X$) that is still open if an update from $X$ has been 
written to \PMN.

\section{Algorithm and Implementation}
\label{sec:impl}

\def \vdb {Volatile Delay Buffer~}

The implementation consists of a user software library backed by a simple \PMCN.
The library is used primarily to coordinate closures of concurrent transactions
with the flow of any data evicted from processor caches into \PM home locations
during those transactions.
The \PMC uses the dependency set concept from
~\cite{spaa17, doshi16, cal16} 
to temporarily park any processor cache eviction in a searchable Volatile Delay Buffer (VDB) so
that its effective time to reach \PM is no earlier than 
the time that the last possible transaction with which the eviction could have
overlapped has become recoverable.
The \PMC in this work improves upon the backend controller of ~\cite{cal16}
by dispensing with synchronous log replays and victim cache management.
The library also covers any writes to \PM variables by volatile write-aside log
entries made within the speculative scope of an HTM transaction; and then streaming
the transactional log record into a non-volatile \PM area outside the HTM
transaction.
These log streaming writes into \PM bypass the VDB.
A software mechanism may periodically check the remaining capacity of the \PM log area and initiate a log cleanup if needed; for such occasional cleanups, new
transactions are delayed, and, after all open transactions have closed, processor
caches are flushed (with a closing sfence), the logs are removed.

We refer to our implementation as {\bf WrAP}, for Write-Aside Persistence, and
individual transactions as wraps.
We first describe the timestamp mechanism, then the user software library, and
finally describe the \PMC implementation details.

\subsection{System Time Stamp}
\label{sec:impl:ts}

We use the recent Intel instruction {\small RDTSCP}, or Read Time Stamp Counter and
Processor ID, to obtain the timestamps in listing 4.
The {\small RDTSCP} instruction provides access to a global monotonically increasing processor
clock across processor 
sockets ~\cite{spear13}, while serializing itself behind the
instructions that precede it in program order.
To prevent the reordering of an {{\small XE}nd} before the {\small RDTSCP} instruction, we save
the resulting time stamp into a volatile memory address.
Since all stores preceding an {{\small XE}nd} become visible after {{\small XE}nd}, and the store
of the persist timestamp is the last store before {{\small XE}nd}, that store gets neither
re-ordered before other stores nor re-ordered after the end of the HTM transaction.
We note that {\small RDTSCP} has also been used to order HTM transactions in novel transaction
profiling ~\cite{tsxprof} and memory version checkpointing ~\cite{ismm17} tools.

\subsection{Software Library}
\label{sec:impl:sw}

\begin{figure}[t]
\centering
\includegraphics[width=0.48\textwidth
]{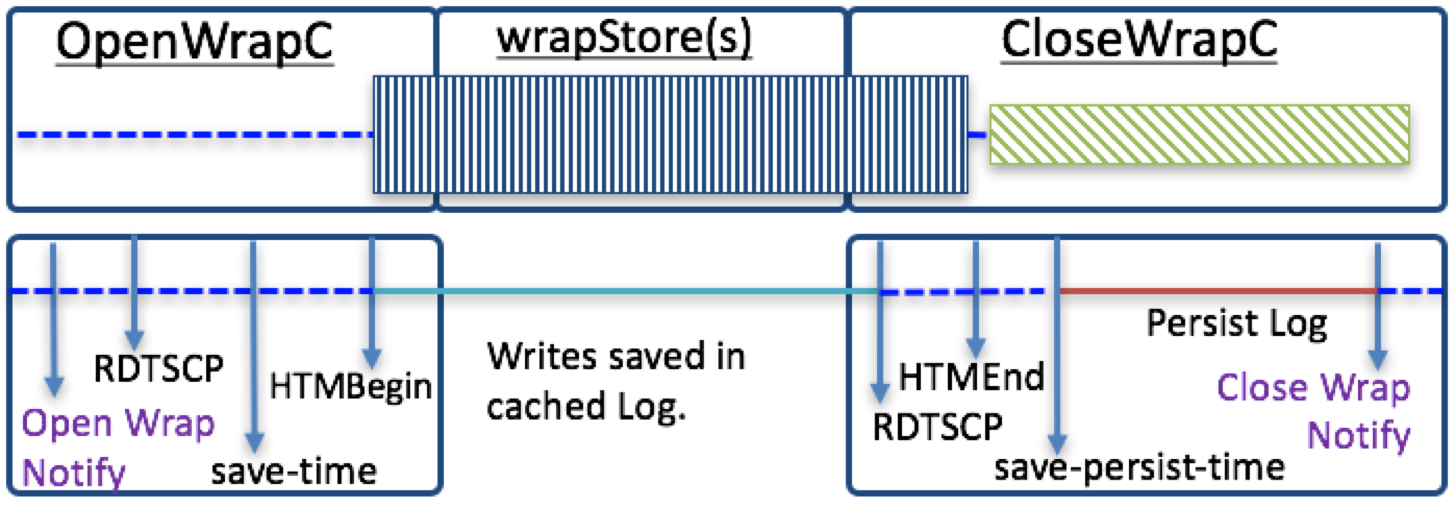}
   \caption{Flow of a transaction with implementation using HTM, cached write sets, timing, and logging.}
   \label{fig:fig-wrap-detail}
\end{figure}

\begin{algorithm}[h]
\caption{Concurrent WrAP User Software Library}
\BlankLine
{\bf User Level Software WrAP Library:}\\
\BlankLine
\BlankLine
\FuncSty{OpenWrapC}
	\ArgSty{()} \\
\indentA {\it // ------------ State: OPEN --------------}\\
\indentA $wrapId$ = threadId\;
\indentA Notify Controller Open Wrap $wrapId$\;
\indentA $startTime$ = RDTSCP()\;
\indentA Log[$wrapId$].startTime = startTime\;
\indentA Log[$wrapId$].writeSet = \{\}\;
\indentA sfence()\;
\indentA CLFLUSH Log[$wrapId$]\;
\indentA sfence()\;
\indentA {\it // ------------ State: COMPUTE ---------}\\
\indentA HTMBegin(); \indent {\it // XBegin}\\
\BlankLine
\BlankLine
\FuncSty{wrapStore}
	\ArgSty{({\bf addrVar}, {\bf Value})} \\
\indentA Add $ \{addrVar, Value\} $ to {\it Log[$wrapId$].writeSet}\;
\indentA Normal store of {\it Value} to {\it addrVar}\;
\BlankLine
\BlankLine
\FuncSty{CloseWrapC}
	\ArgSty{({\bf strictDurability})}\\
\indentA Log[$wrapId$].persistTime = RDTSCP()\;
\indentA HTMEnd(); \indent {\it // XEnd}\\
\indentA {\it // ------------ State: LOG ------------------}\\
\indentA CLFLUSH Log[$wrapId$].persistTime\;
\indentA for num cachelines in Log[$wrapId$].writeSet\\
\indentB CLFLUSH cacheline\;
\indentA if ($strictDurability$) \\
\indentB $durabilityAddr$ = 0; // Reset Flag\;
\indentB $tAddr$ = $durabilityAddr$\;
\indentA else $tAddr$ = 0\;
\indentA sfence()\;
\indentA {\it // ------------ State: CLOSE ---------------}\\
\indentA Notify Controller Closed ($wrapId$, $tAddr$)\;
\indentA {\it // ------------ State: COMMIT ------------}\\
\indentA if ($strictDurability$)\\
\indentB //  Wait for durable notification from controller\\
\indentB Monitor $durabilityAddr$\;
\label{algo:SwWrapAlgo}
\vspace{0pt}
\end{algorithm}

For HTM we employ Intel's implementation of Restricted Transactional Memory or
RTM, which includes the instructions XBegin and XEnd.
Aborting HTM transactions retry with exponential back-off a few times, and then
are performed under a software lock.
Our HTMBegin routine checks the status of the software lock both before and after
an XBegin, to produce the correct indication of conflicts with the non-speculative
paths; acquiring the software lock non-speculatively after having backed off.
HTMBegin and HTMEnd library routines perform the acquire and release of the software
lock for the fallback case within themselves.
The remaining software library procedures are shown in 
Algorithm ~\ref{algo:SwWrapAlgo}.
Various events that arise in the course of a transaction are shown in 
Figure ~\ref{fig:fig-wrap-detail},
which depicts the HTM concurrency section with vertical lines and the
logging section with slanted lines.

Not shown in 
Figure ~\ref{fig:fig-wrap-detail},
is a per-thread {\it durability address} location that we call {\it durabilityAddr} in 
Algorithm ~\ref{algo:SwWrapAlgo}.
A software thread may use it to setup a Monitor-Mwait coordination to be
signaled via memory by the \PMC (as described shortly) when a transacting
thread wants to wait until all updates from any non-conflicting transactions that
may have raced with it are confirmed to be recoverable.

This provision allows for implementing the strict durability for any transaction
because the logs of all other transactions that could possibly precede it in
persistence order are in \PM --which guarantees the replayability of its log.
By contrast, many other transactions that only need the consistency guarantee
(correct log ordering) may continue without waiting (or defer waiting to a
different point in the course of a higher level multi-transaction operation).
The number of active HTM transactions at any given time is bounded by the number
of CPUs, therefore, we use thread identifiers as wrapIds.
In {\bf OpenWrapC} we notify the \PMC that a wrap has started.
We then read the start time with {\small RDTSCP} and save it and an empty write set into
its log persistently.
The transaction is then started with the HTMBegin routine.

During the course of a transactional computation, the stores are performed using
the {\bf wrapStore} function.
The stores are just the ordinary (speculatively performed) store instructions,
but are accompanied by (speculative) recording of the updates into the log
locations, each capturing the address, value pair for each 
update, to be committed into \PM later during the logging phase (after XEnd).

In {\bf CloseWrapC} we obtain and record the ending timestamp for an HTM transaction
into the persistTime variable in its log.
Its concurrency section is then terminated with the HTMEnd routine.
At this point, the cached write set for the log and ending persistent timestamp
are instantly visible in the cache.
Next, we flush transactional values and the persist timestamp to the log area
followed by a persistent memory fence.
The transaction closure is then notified to the \PMC with the wrapId,
and along with it, the durabilityAddr, if the thread has requested $strict$
durability (by passing a flag to {\bf CloseWrapC}) -- for which, we 
use the efficient Monitor-Mwait construct to receive memory based
signaling from the \PMCN.
If strict durability is not requested, then CloseWrapC can return immediately
and let the thread proceed immediately with $relaxed$ durability.
In many cases a thread performing a series of transactions may choose relaxed
durability over all but the last and then request strict durability over the entire
set by waiting for only the last one to be strictly durable.

\subsection {PM Controller Implementation}
\label{sec:impl:hw}

\begin{algorithm}[t!]
\caption{Hardware WrAP Implementation}
\BlankLine
{\bf \PMC:}\\
\BlankLine
\BlankLine
\FuncSty{Open Wrap Notification}
	\ArgSty{({\bf wrapId})} \\
\indentA Add $wrapId$ to Current Open Transactions COT\;
%
\BlankLine
\BlankLine
\FuncSty{Memory Write}
\ArgSty{({\bf memoryAddr, data})}\\
\indentA //  Triggered from cache evict or stream store.\\
\indentA if (($memoryAddr$ not in Pass-Through Log Area) and \\
\indentA \indent (Current Open Transactions COT != \{\}))\\
\indentB Add ($memoryAddr$, $data$, $COT$) to VDB\;
\indentA else\\
\indentB //  Normal memory write\\
\indentB Memory[$memoryAddr$] = $data$\;
\BlankLine
\BlankLine
\FuncSty{Memory Read}
\ArgSty{({\bf memoryAddr})}\\
\indentA if ($memoryAddr$ in Volatile Delay Buffer)\\
\indentB return latest cacheline data from VDB\;
\indentA return Memory[$memoryAddr$]\;
\BlankLine
\BlankLine
\FuncSty{Close WrAP Notification}
\ArgSty{({\bf wrapId, durabilityAddr})}\\
\indentA Remove $wrapId$ from Current Open Transactions COT\;
\indentA if ($durabilityAddr$)\\
\indentB $durabilityDS$ = COT\;
\indentB //  Add pair to Durability Wait Queue DWQ\; 
\indentB Add ($durabilityDS$, $durabilityAddr$) to DWQ\;
\indentA Remove $wrapId$ from Volatile Delay Buffer elements\\
\indentA if earliest VDB elements have empty DS\\
\indentB //  Write back entries to memory in FIFO order\\
\indentB Memory[$memoryAddr$] = data\;
\indentA Remove $wrapId$ from Durability Wait Queue elements\\
\indentA if earliest DWQ elements have empty DS\\
\indentB //  Notify waiting thread of durability\\
\indentB Memory[$durabilityAddr$] = 1\;
\label{algo:HwWrapAlgo}
\vspace{0pt}
\end{algorithm}

The \PMC provides for two needs: (1) holding back modified \PM cachelines that fall into it at any time T from the processor caches, from
flowing into \PM until at least a time when all successful transactions that were active at
time T are recoverable, and (2) tracking the ordering of dependencies among
transactions so that only those that need strict durability guarantees need to
be delayed pending the completion of the log phases of those with which
they overlap.
It implements a VDB (volatile data buffer) as means for the transient storage
for the first need, implements a durability wait queue (DWQ) for the second
need, and implements a dependency set (DS) tracking logic across the
open/close notifications to control the VDB and the DWQ, as described next.

{\bf Volatile Delay Buffer ({\small VDB}):~~}
The {\small VDB} is comprised of a {\small FIFO} queue and hash table that points to entries in the {\small FIFO} queue.
Each entry in the {\small FIFO} queue contains a tuple of {\it \PM address, data,} and {\it dependency set}.
On a \PM write, resulting from a cache eviction or streaming store, to a memory address not in the log area or pass-through area, the \PM address and data are added to the {\small FIFO} queue and tagged with a dependency set initialized to the {\small COT}.
Additionally, the \PM address is inserted into the hash table with a pointer to the {\small FIFO} queue entry.
If the address already exists in the hash table, then it is updated to point to the new queue entry.
On a memory read, the hash table is first consulted.
If an entry is in the hash table, then the pointer is to the latest memory value for the address, and the data is retrieved from the queue.
On a hash table miss, \PM is read and data is returned.
As wraps close, the dependency set in each entry in the queue is updated to remove the dependency on the wrap.

Dependency sets become empty in {\small FIFO} order, and as they become empty, we perform three actions.
First, we write back the data to the \PM address.
Next, we consult the hash table.
If the hash table entry points to the current {\small FIFO} queue entry, we remove the entry in the hash table, since we know there are no later entries for the same memory address in the queue.
Finally, we remove the entry from the {\small FIFO} queue.

On inserting an entry into the back of the queue, we can also consult the head of the {\small FIFO} queue to check to see if the dependency set is empty.
If the head has an empty dependency set, we can perform the same actions, allowing for O(1) {\small VDB} management.

{\bf Dependency Wait Queue ({\small DWQ}):~~}
Strict durability is handled by the \PMC using the Dependency Wait Queue or {\small DWQ}, which is used to track transactions waiting on others to complete and notify the transaction that it is safe to proceed.
The {\small DWQ} is a {\small FIFO} queue similar to the {\small VDB} with entries containing pairs of the {\it dependency set} and a {\it durability address}.

When a thread notifies the \PMC that it is closing a transaction (see steps below), it can request $strict$ durability by passing a $durability~address$.
Dependencies on closing wraps are also removed from the $dependency~set$ for each entry in the {\small DWQ}.
When the $dependency~set$ becomes empty, the controller writes to the $durability~address$ and removes the entry from the queue.
Threads waiting on a write to the address can then proceed.

{\bf Opening and Closing WrAPs:~~}
As outlined in 
Algorithm ~\ref{algo:HwWrapAlgo},
the controller supports two interfaces from
software, namely those for {\it Open Wrap} and {\it Close Wrap} notifications exercised
from the user library as shown in Algorithm ~\ref{algo:SwWrapAlgo}.
(Implementations of these notification can vary: for example, one possible
mechanism may consist of software writing to a designated set of control
addresses for these notifications).
It also implements hardware operations against the VDB from the processor caches:
Memory Write, for handling modified cachelines evicted from the processor caches
or non-temporal stores from CPUs and Memory Read, for handling reads from \PM from the processor caches.

The Open Wrap notification simply adds the passed (wrapId) to a bit vector of open transactions.
We call this bit vector of open transactions the Current Open Transactions COT.
When the controller receives a Memory Write (i.e., a processor cache eviction or a non-temporal/streaming/uncached write) it checks the COT: if the COT is empty, writes can flow into the \PMN.
Writes that target the log range in \PM can also flow into \PM irrespective of the COT.
For the non-log writes, if the COT is nonempty cache line is tagged with the COT and placed into the VDB.

The {\bf Close Wrap} controller notification receives the {\it wrapId} and durability address, $durabilityAddr$.
The controller removes the $wrapId$ from the Current Open Transactions {\small COT} bit mask.
If the transaction requires $strict$ durability, we save the $durabilityDS$ and {\small COT} as a pair in the {\small DWQ}.
The controller then removes the {\it wrapId} from all entries in the {\small VDB} and {\small DWQ}.
This is performed by simply draining the bit on the dependency set bit mask for the entire {\small FIFO VDB}.
If the earliest entries in the queue result in an empty dependency set, the cache line data is written back in {\small FIFO} order.
Similarly, the controller removes the $wrapId$ from all entries in the Durability Wait Queue {\small DWQ}.

{\bf Software Based Strict Durability Alternative:}
As an alternative for implementing strict durability in the controller, strict durability may be implemented entirely in the software library; we modify the software algorithm as follows.
On a transaction start, threads save the start time and an open flag in a dedicated cache line for the thread.
On transaction close, to ensure strict durability, it saves its end time in the same cache line with the start time and clears the open flag.
It then waits until all prior open transactions have closed.
It scans the set of all thread cache lines and compares any open transaction end times and start times to its end time.
The thread may only continue, with ensured durability, once all other threads are either not in an open transaction or have a start or persist time greater than its persist time.

\begin{figure*}[t!]
\centering
\begin{subfigure}[b]{0.45\textwidth}
        \includegraphics[width=\textwidth,height=2.3in]{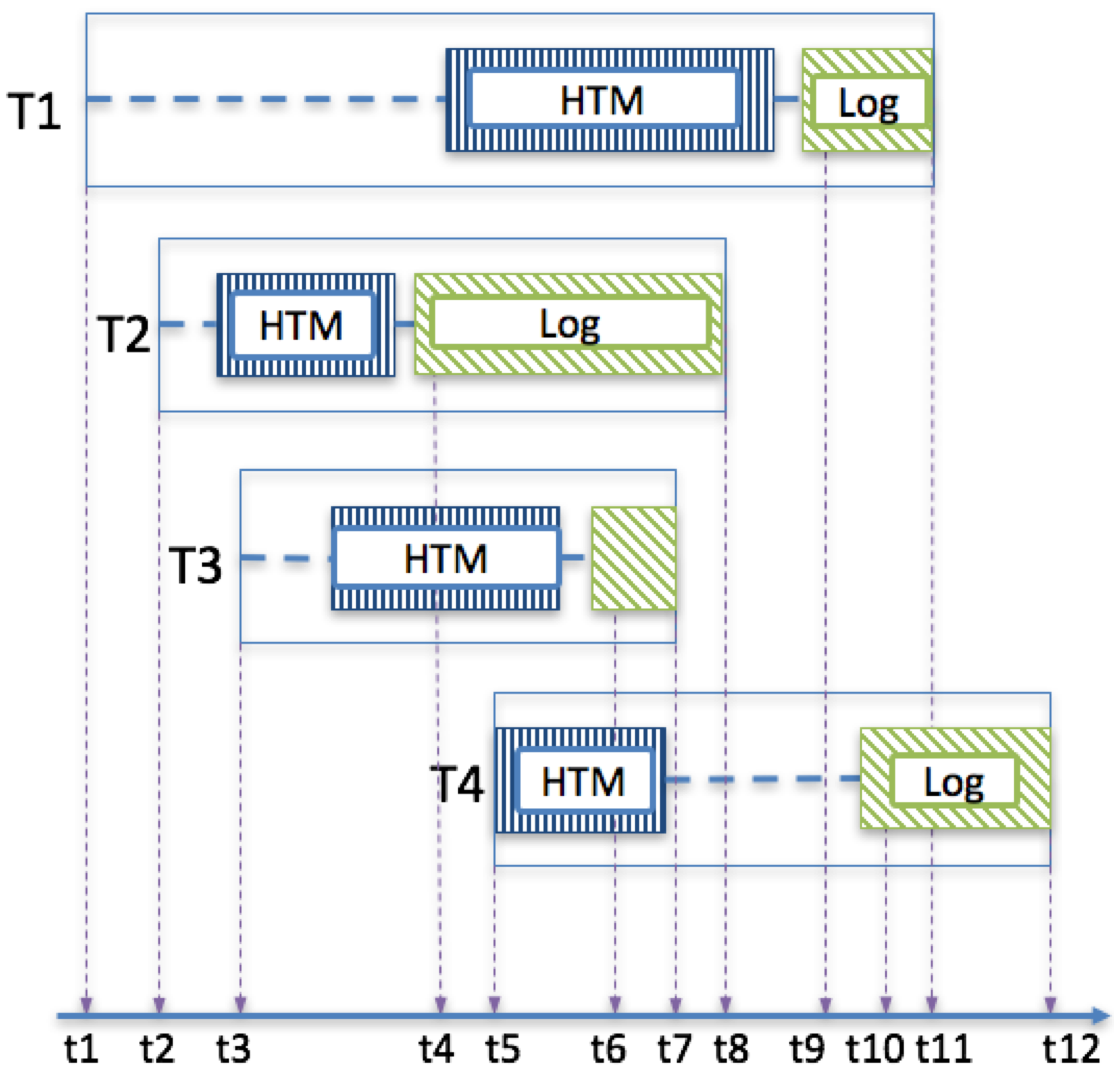}
        \caption{Example Transactions T1-T4}
        \label{fig:example}
    \end{subfigure}
\hfill
\begin{subfigure}[b]{0.45\textwidth}
\centering
\begin{tabular}{|l|l|l|}
\hline
\textbf{\small Time}\rule{0pt}{2.9ex}\rule[-1.2ex]{0pt}{0pt} & { \small \textbf{\begin{tabular}{@{}c@{}}Only Start Timestamp\\ Has Been Persisted \end{tabular}}} & 
			{ \small \textbf{\begin{tabular}{@{}c@{}}Order of Persist\\ Timestamps \end{tabular}}} \\  \hline
\textbf{t1}   & { T1}                  & { }                   \\ \hline
\textbf{t2}   & { T1, T2}              & { }                   \\ \hline
\textbf{t3}   & { T1, T2, T3}          & { }                   \\ \hline
\textbf{t4}   & { T1, T3}              & { T2}                 \\ \hline
\textbf{t5}   & { T1, T3, T4}          & { T2}                 \\ \hline
\textbf{t6}   & { T1, T4}              & { T2, T3}             \\ \hline
\textbf{t7}   & { T1, T4}              & { T2, {\bf \underline{T3}}}             \\ \hline
\textbf{t8}   & { T1, T4}              & { {\bf \underline{T2}}, {\bf \underline{T3}}}             \\ \hline
\textbf{t9}   & { T4}                  & { {\bf \ovaled{T2}}, {\bf \underline{T3}}, T1}         \\ \hline
\textbf{t10}  & { }                    & { {\bf \ovaled{T2}}, {\bf \ovaled{T3}}, T4, T1}         \\ \hline
\textbf{t11}  & { }                    & { {\bf \ovaled{T2}}, {\bf \ovaled{T3}}, T4, {\bf \underline{T1}}}             \\ \hline
\textbf{t12}  & { }                    & { {\bf \ovaled{T2}}, {\bf \ovaled{T3}}, {\bf \ovaled{T4}}, {\bf \ovaled{T1}}}             \\ \hline
\end{tabular}
\caption{Transaction Recovery Order on Failure}
\label{table:recover}
\end{subfigure}

\vspace{4pt}
\begin{subfigure}[b]{\textwidth}
\centering
\resizebox{\textwidth}{!}{%
\begin{tabular}{|l|l|l|l|l|l|l|l|l|l|l|}
\hline
\textbf{Time}       & \textbf{t1 - t3} & \textbf{t4}    & \textbf{t5}    & \textbf{t6}                                                             & \textbf{t7}                                                            & \textbf{t8}                                                             & \textbf{t9}                                                                              & \textbf{t10}                                                                                              & \textbf{t11}                                                                                              & \textbf{t12}                                                                             \\ \hline
\textbf{Event}      & T1,T2,T3 Start   & Evict X        & T4 Starts      & Evict Y                                                                 & T3 Ends                                                                & T2 Ends                                                                 & Evict Z                                                                                  & Evict X                                                                                                   & T1 Ends                                                                                                   & T4 Ends                                                                                  \\ \hline
\textbf{\small COT}    & \{1,1,1,0\} & \{1,1,1,0\}    & \{1,1,1,1\}    & \{1,1,1,1\}                                                             & \{1,1,0,1\}                                                            & \{1,0,0,1\}                                                             & \{1,0,0,1\}                                                                              & \{1,0,0,1\}                                                                                               & \{0,0,0,1\}                                                                                               & \{0,0,0,0\}                                                                              \\ \hline
\textbf{\begin{tabular}{@{}c@{}} $\downarrow$ \\ Volatile \\ Delay \\ Buffer \\ $\downarrow$ \\ \end{tabular}} & -           & 
		\begin{tabular}[c]{@{}l@{}} $\downarrow$\\ X: \{1,1,1,0\} \\ \\ \\ \\ \end{tabular} & 
		\begin{tabular}[c]{@{}l@{}}  \\ X: \{1,1,1,0\} \\ \\ \\ \\ \end{tabular} & 
		\begin{tabular}[c]{@{}l@{}} $\downarrow$ \\ Y: \{1,1,1,1\}\\ X: \{1,1,1,0\} \\ \\ \\ \end{tabular} & 
		\begin{tabular}[c]{@{}l@{}} \\ Y: \{1,1,0,1\}\\ X: (1,1,0,0\} \\ \\ \\ \end{tabular} & 
		\begin{tabular}[c]{@{}l@{}} \\ Y: \{1,0,0,1\}\\ X: \{1,0,0,0\} \\ \\ \\ \end{tabular} & 
		\begin{tabular}[c]{@{}l@{}} $\downarrow$ \\ Z: \{1,0,0,1\}\\ Y: \{1,0,0,1\}\\ X: \{1,0,0,0\} \\ \\ \end{tabular} & 
		\begin{tabular}[c]{@{}l@{}} $\downarrow$ \\ X: \{1,0,0,1\}\\ Z: \{1,0,0,1\}\\ Y: \{1,0,0,1\}\\ X: \{1,0,0,0\}\end{tabular} & 
		\begin{tabular}[c]{@{}l@{}}X: \{0,0,0,1\}\\ Z: \{0,0,0,1\}\\ Y: \{0,0,0,1\}\\ X: \{0,0,0,0\} \\ $\downarrow$ \end{tabular} & 
		\begin{tabular}[c]{@{}l@{}} \\ X: \{0,0,0,0\}\\ Z: \{0,0,0,0\}\\ Y: \{0,0,0,0\} \\ $\downarrow$ \end{tabular} \\ \hline
\end{tabular}
}
\caption{Contents of the Persistent Memory Controller Current Open Transactions Dependency Set and FIFO Queue}
\label{table:nvm}
\end{subfigure}
\vspace{-0pt}
\caption{Example Transaction Sequence with Contents of Persistent Memory Controller and Recovery Algorithm}
\vspace{-0pt}
\end{figure*}

\def \cds {{\small COT}~}
\def \currds {Current Open Transactions~}

\subsection{Example}
\label{sec:example}

Figure ~\ref{fig:example} shows an example set of four transactions, T1-T4, happening concurrently.
In this example, we show T1-T4 split into states, specifically, concurrency ({\small COMPUTE}), shown in vertical lines, and {\small LOG}, depicted with slanted lines.
At certain time steps we show the contents of the \PMCN's Volatile Delay Buffer, or {\small VDB} which is a {\small FIFO} Queue, and the \currds or \cds in Figure ~\ref{table:nvm}.
The recovery algorithm is shown in Figure ~\ref{table:recover} with the contents of the log.
Either the start timestamp only or the persist timestamp order is shown; where bold and underline indicate the log is written, and a circled transaction indicates that it is recoverable.

First, at time $t1$, T1 opens, notifying the controller, and records its start timestamp safely in its log.
The controller adds T1 to the bitmap \cds of open transactions.
At times $t2$ and $t3$, transactions T2 and T3 also open, notify the controller, and safely read and persist their start timestamps.
At this point in time the \PMC has a \cds of \{1,1,1,0\} and only start timestamps have been recorded in the log.
T2 then completes its concurrency section and persists its persist timestamp at time $t4$ and begins writing its log.
In Figure ~\ref{table:nvm}, we also show a random cache eviction of a cache line $X$ that is tagged with the \cds of \{1,1,1,0\}.

Transaction T4 starts at time $t5$ persisting its start timestamp and is added to the set of open transactions on the \PMCN, now \{1,1,1,1\}.
At time $t6$ we illustrate several events.
T3 completes its concurrency section and persists its persist timestamp.
At this time, as shown in ~\ref{table:recover}, T3 is now ordered after T2 for recovery, however neither have completed persisting their logs.
Also, we show a random cache eviction of cache line $Y$, and it is placed at the back of the {\small VDB} on the \PMC as shown in ~\ref{table:nvm}.

At time $t7$, transaction T3 has completed writing its logs and is marked completed and is removed from the dependency set in the controller and cache line dependencies for $X$ and $Y$.
However, as shown in ~\ref{table:recover}, T3 is not recoverable at this point since it is not first in the persist timestamp order.
T3 is behind T2 and T3 also has a smaller persist timestamp than the start timestamp of T1, which hasn't written its persist timestamp yet.
When T2 completes writing of logs at time $t8$, it is removed from the current and dependency sets in the queue of the \PMC as shown in ~\ref{table:nvm}.
Note that cache line $X$ is still not able to be written back to \PMemory as it is still tagged as being dependent on T1, and $Y$ on both T1 and T4.
At this time, T2 and T3 are also not recoverable as shown in ~\ref{table:recover} as T2 has a persist timestamp that is greater than the start timestamp of T1, as it would be unknown by a recovery process if T1 had simply had a delay in persisting its log and T2 had transactional values dependent on T1.

We illustrate two events at time $t9$.
T1 finally completes its concurrency section and writes its persist timestamp at $t9$.
Since the persist timestamp of T1 is safely persisted and known at recovery time, transaction T2 is now fully recoverable as shown circled in ~\ref{table:recover}.
However, at this time, T3 is not fully recoverable since it is waiting on T4, which started before T3 completed its concurrency section and T4 hasn't yet written the persist timestamp.
Also at $t9$, in ~\ref{table:nvm} we illustrate the eviction of cache line $Z$ which is tagged with the set of open transactions {\small COT} \{1,0,0,1\}.

At time $t10$, T4 writes its persist timestamp and its order is now known to a recovery routine to be behind T3, which is now fully recoverable as shown with the circle in ~\ref{table:recover}.
Note that T4 has a persist time before T1.
In ~\ref{table:nvm}, we also illustrate the eviction of cache line $X$ again into the {\small VDB} of the \PMC and tagged with the set of the two open transactions, T1 and T4.
Note that there are two copies of cache line $X$ in the controller.
The one at the head of the queue has fewer dependencies (only dependent on T1) than the recent eviction.
Any subsequent read for cache line $X$ returns the most recent copy, the last entry in the {\small VDB}.
Note how cache lines at the back of the queue have dependency set sizes that are greater than or equal to entries earlier in the queue.

T1 completes log writing at $t11$, but is behind T4, which hasn't yet finished writing its logs, so neither are yet recoverable.
The \PM controller also removes T1 from its dependency set and of those in the {\small VDB}.
The first copy of $X$ now has no dependencies in the queue and is safely written back to \PM as shown in ~\ref{table:nvm}.

At time $t12$, T4 completes writing its logs and both T4 and T1 are recoverable.
Also, T4 is removed from the dependency sets in the controller, which allows for $Y$, $Z$, and $X$ to flow to \PM.

{\bf Strict Durability:~}
Suppose a transaction requires $strict$ durability during its Commit Stage, ensuring that once complete, the transactional writes will be reflected in \PM if a failure were to occur.
If T4 requires $strict$ durability, it is simply durable at the end as there are no open transactions when it completes.
However, T1, T2, and T3, have other constraints.
A transaction requiring strict durability is only durable when it is fully recoverable.
Table ~\ref{table:recover} illustrates transaction durability when it is circled.
T1 must wait until step $t12$ if it requires strict durability as it might have dependencies on T4.
T2 is fully durable at time $t9$ when T1, which started earlier, writes its persist timestamp.
At time $t10$, T3 is fully durable when T4, which started before T3 completed its concurrency section and could have introduced transactional dependencies, writes its persist timestamp which indicates T4 started {\small HTM} section later.

\section{Evaluation}
\def\plotwidth{3.05in}
\def\plotheight{2.44in}

We evaluated our method using benchmarks directly running on hardware and through simulation analysis (described in Section ~\ref{sec:simulation}).
Our simulation evaluates the length of the {\small FIFO} buffer and performance against various \PMemory write times.
In the direct hardware evaluation described next, we employed Intel(R) Xeon(R) E5-2650 v4 series processors, 12 cores per processor, running at 2.20 GHz, with Red Hat Enterprise Linux 7.2.
{\small HTM} transactions were implemented with Intel Transactional Synchronization Extensions ({\small TSX}) ~\cite{IntelTSX} using a global fallback lock.
We built our software using g++ 4.8.5.
Each measurement reflects an average over twenty repeats with small variation among the repeats.

Using micro-benchmarks and {\small SSCA2}~\cite{ssca} and {\it Vacation}, from the {\small STAMP} ~\cite{stamp} benchmark suite, we compared the following methods:
\begin{itemize}
\item {\bf {\small HTM} Only:}~Hardware Transactional Memory with Intel {\small TSX}, without any logging or persistence.
This method provides a baseline for transaction performance in cache memory without any persistence guarantees.
If a power failure occurs after a transaction, writes to memory locations may be left in the cache, or written back to memory in an out-of-order subset of the transactional updates.
\item {\bf {\small WrAP}:}~Our method.  The volatile delay buffer in the controller is assumed to be able to keep up with back pressure from the cache, as shown in Section 3.  We therefore perform all other aspects of the protocol such as logging, reading timestamps, {\small HTM} and fall-back locking, etc.
\item {\bf {\small WrAP-Strict}:}~Same as above, but we implement the software strict durability method as described in Section 4.  Threads wait until all prior-open transactions have closed before proceeding.
\item {\bf {\small PTL-Eager}:}~(Persistent Transactional Locking).
In this method, we added persistence to Transactional Locking ({\small TL}-Eager)~\cite{TL,TL2, tl2x86} by persisting the undo log at the time that a {\small TL} transaction performs its sequence of writes. The undo-log entries are written with write-through stores and {\small SFENCE}s, and once the transaction commits and the new data values are flushed into \PMN, the undo-log entries are removed.
\end{itemize}

\subsection{Benchmarks}

The Scalable Synthetic Compact Applications for benchmarking High Productivity Computing Systems ~\cite{ssca}, {\small SSCA2}, is part of the Stanford Transactional Applications for Multi-Processing ~\cite{stamp}, or {\small STAMP}, benchmark suite.
{\small SSCA2} uses a large memory area and has multiple kernels that construct a graph and perform operations on the graph.
We executed the {\small SSCA2} benchmark with scale 20, which generates a graph with over 45 million edges.
We increased the number of threads from 1 to 16 in powers of two and recorded the execution time for the kernel for each method.

\begin{figure}[th]
\centering
\includegraphics[width=\plotwidth,height=\plotheight]{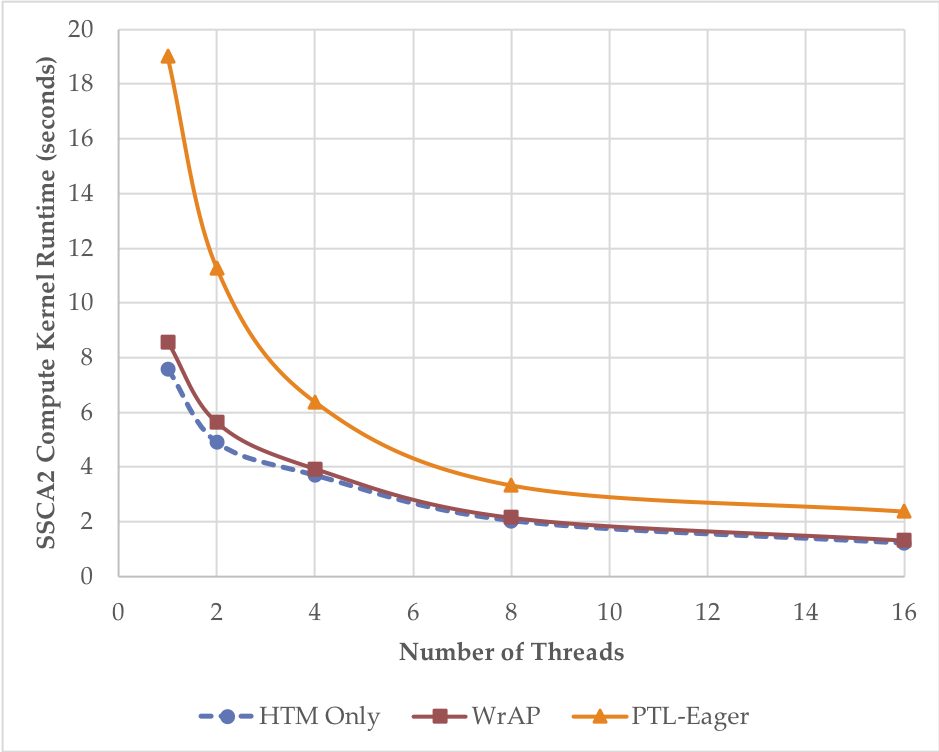}
\caption{{\small SSCA2} Benchmark Compute Graph Kernel Execution Time as a Function of the Number of Parallel Execution Threads}
\label{fig:s30}
\end{figure}

\begin{figure}[th]
\centering
\includegraphics[width=\plotwidth,height=\plotheight]{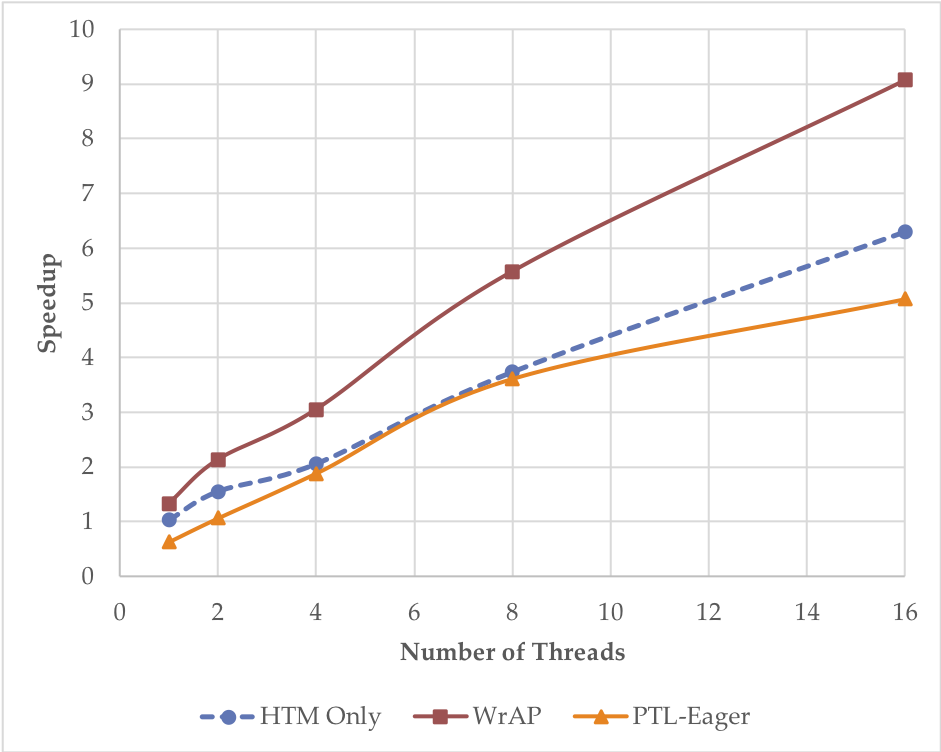}
\caption{{\small SSCA2} Benchmark Compute Graph Kernel Speedup as a Function of the Number of Parallel Execution Threads}
\label{fig:s31}
\end{figure}

\begin{figure}[th]
\centering
\includegraphics[width=\plotwidth,height=\plotheight]{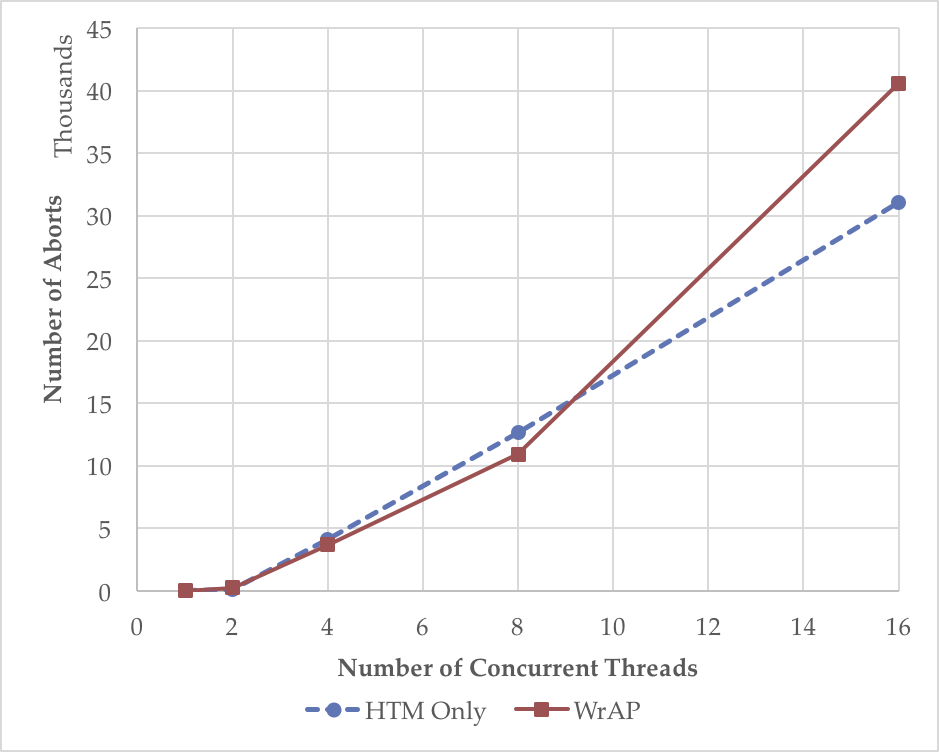}
\caption{{\small SSCA2} Benchmark Compute Graph Kernel {\small HTM} Aborts as a Function of the Number of Parallel Execution Threads}
\label{fig:s32}
\end{figure}

Figure ~\ref{fig:s30} shows the execution time for each method for the Compute Kernel in the {\small SSCA2} benchmark as a function of the number of threads.
Each method reduces the execution time with increasing numbers of threads.
Our WrAP approach has similar execution time to {\small HTM} in the cache hierarchy with no persistence and is over 2.25 times faster than a persistent {\small PTL-Eager} method to \PMN.

Figure ~\ref{fig:s31} shows the speedup for each method as a function of the number of threads when compared to a single threaded undo log for the persistence methods and speedup versus no persistence for the in cache method {\small HTM} only.
Even though the {\small HTM} (cache-only) method does better in absolute terms as we saw in Figure ~\ref{fig:s30}, it proceeds from a higher baseline for single-threaded execution.
{\small PTL-Eager} yields a significantly weaker scalability due to the inherent costs of having to perform persistent flushes within its concurrent region.

Figure ~\ref{fig:s32} shows the number of hardware aborts for both our WrAP approach and cache-only {\small HTM}.
Our approach introduces extra writes to log the write-set, and, along with reading the system time stamp, extends the transaction time.
However, as shown in the Figure, this only slightly increases the number of hardware aborts.

\begin{figure}[t]
\centering
\includegraphics[width=\plotwidth,height=\plotheight]{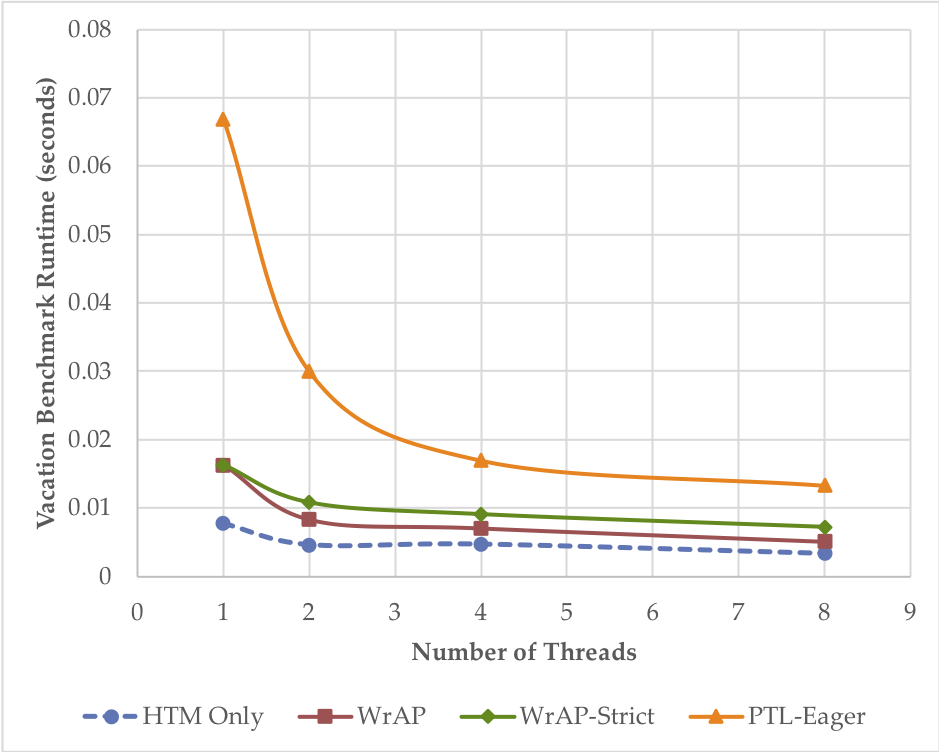}
\caption{Vacation Benchmark Execution Time as a Function of the Number of Parallel Execution Threads}
\label{fig:s33}
\end{figure}

\begin{figure}[t]
\centering
\includegraphics[width=\plotwidth,height=\plotheight]{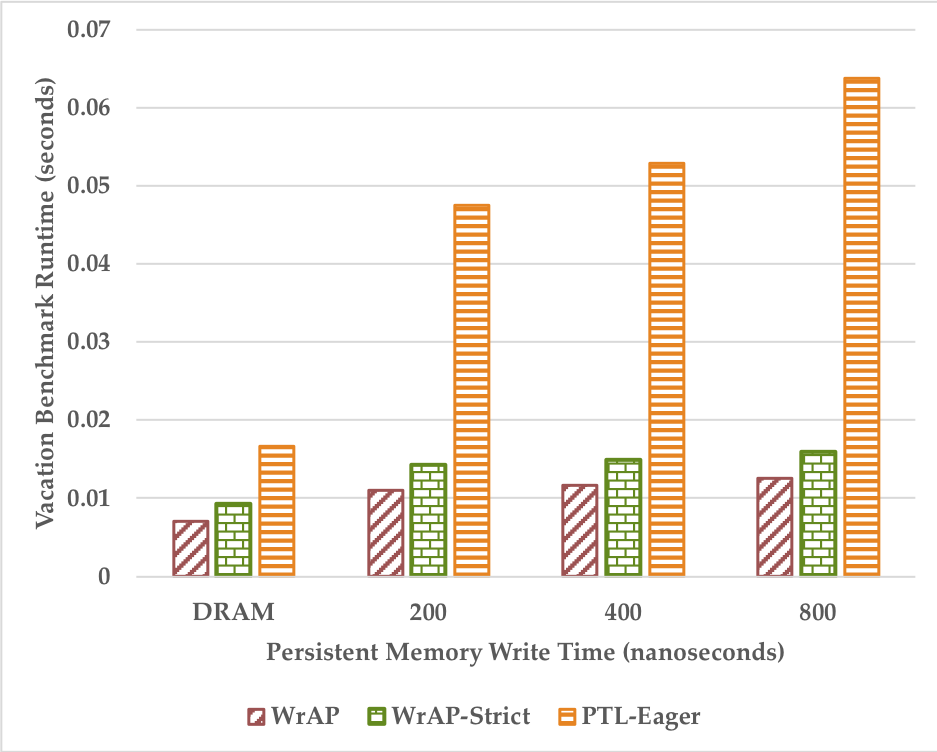}
\caption{Vacation Benchmark Execution Time with Various Persistent Memory Write Times for Four Threads}
\label{fig:s34}
\end{figure}

We also evaluated the {\it Vacation} benchmark which is part of the {\small STAMP} benchmark suite.
The {\it Vacation} benchmark emulates database transactions for a travel reservation system.
We executed the benchmark with the low option for lower contention emulation.
Figure ~\ref{fig:s33} shows the execution time for each method for the {\it Vacation} benchmark as a function of the number of threads.
Each method reduces the execution time with increasing numbers of threads.
The WrAP approach follows the trends similar to {\small HTM} in the cache hierarchy with no persistence, with both approaches flattening execution time after 4 threads.
We examine the effect of $strict$ durability, WrAP-Strict in the figure, and show that $strict$ durability only introduces a small amount of overhead.
For just a single thread, it has the same performance as WrAP relaxed as a thread doesn't need to wait on other threads, as it is durable as soon as the transaction completes.

Additionally, we examined the effect of increased \PMemory write times on the benchmark.
Byte-addressable Persistent Memory can have longer write times.
To emulate the longer write times for \PMN, we insert a delay after non-temporal stores when writing to new cache lines and a delay after cache line flushes.
The write delay can be tuned to emulate the effect of longer write times typical of \PMN.
Figure ~\ref{fig:s34} shows the {\it Vacation} benchmark execution time for various \PM write times.

The WrAP approach is less affected by increasing \PM write times than the {\small PTL-Eager} approach due to several factors.
WrAP performs write-combining for log entries on the foreground path for each thread, so writes to several transaction variables may be combined into a fewer writes.
Also, {\small PTL-Eager} transactionally persists an undo log on writes causing a foreground delay.

\subsection{Hash Table}

Our next series of experiments show transaction sizes and high memory traffic affect overall performance.
We create a 64 MB Hash Table Array of elements in main memory and transactionally perform a number of element updates.
For each transaction, we generate a set of random numbers of a configurable size, compute their hash, and write the value into the Hash Table Array.

\begin{figure}[th]
\centering
\includegraphics[width=\plotwidth,height=\plotheight]{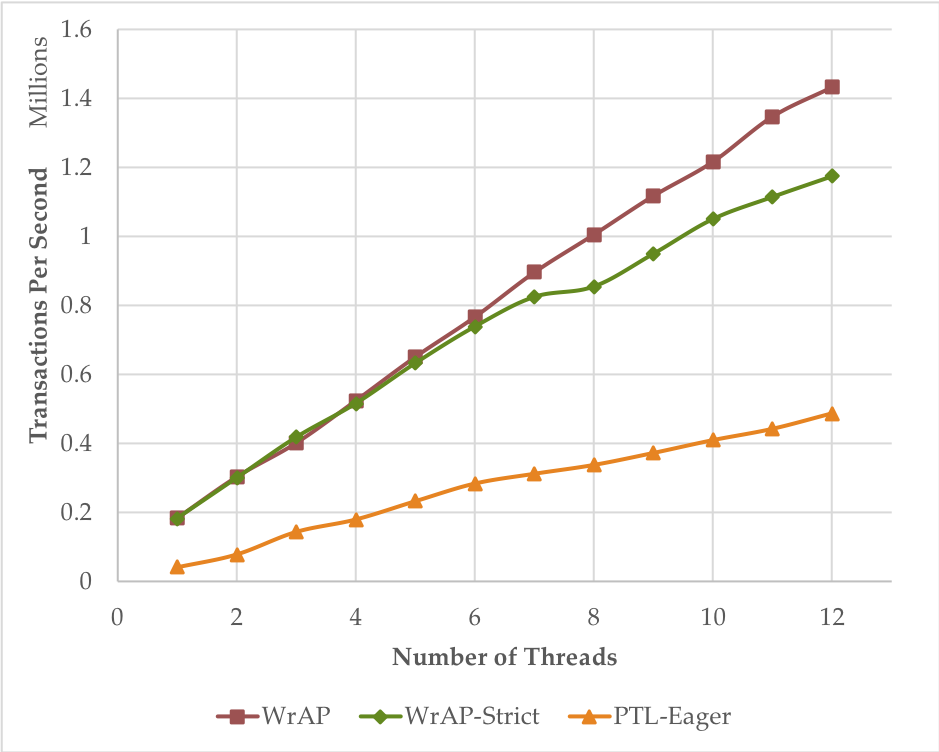}
\caption{Millions of Transactions per Second for Hash Table Updates of 10 Elements versus Concurrent Number of Threads}
\label{fig:s35}
\end{figure}

\begin{figure}[th]
\centering
\includegraphics[width=\plotwidth,height=\plotheight]{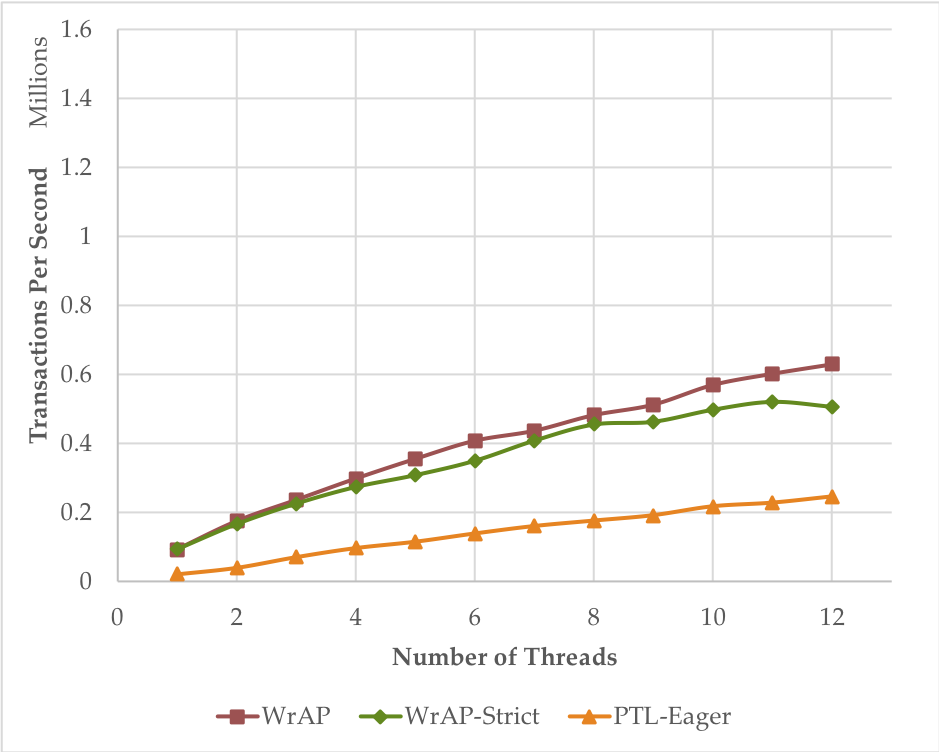}
\caption{Millions of Transactions per Second for Hash Table Updates of 20 Elements versus Concurrent Number of Threads}
\label{fig:s35b}
\end{figure}

First, we create transactions consisting of 10 atomic updates and vary the number of concurrent threads and measure the maximum throughput.
We perform 1 million updates and record the average throughput and plot the results in Figure ~\ref{fig:s35}.
Our approach achieves roughly 3x throughput over {\small PTL-Eager}.
Figure ~\ref{fig:s35b} shows increasing the write set to 20 atomic updates has similar performance.
In both figures, adding $strict$ durability only slightly decreases the overall performance; threads wait additional time for the dependency on other transactions to clear before continuing to another transaction.

\begin{figure}[th]
\centering
\includegraphics[width=\plotwidth,height=\plotheight]{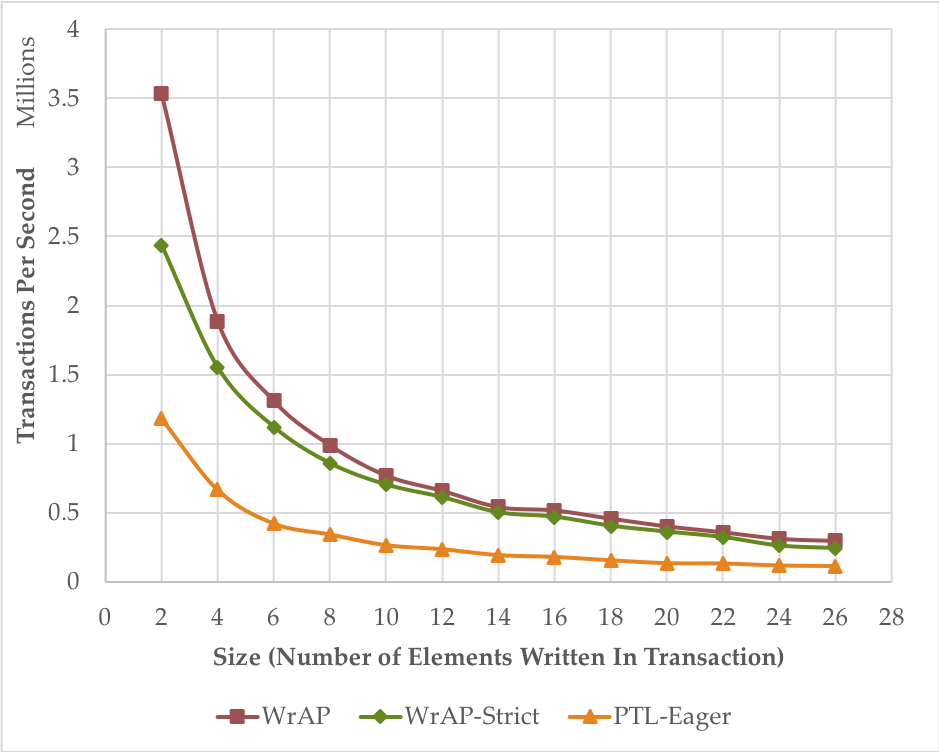}
\caption{Average Txps for Increasing Transaction Sizes of Atomic Hash Table Updates with 6 Concurrent Threads}
\label{fig:s36}
\end{figure}

\begin{figure}[th]
\centering
\includegraphics[width=\plotwidth,height=\plotheight]{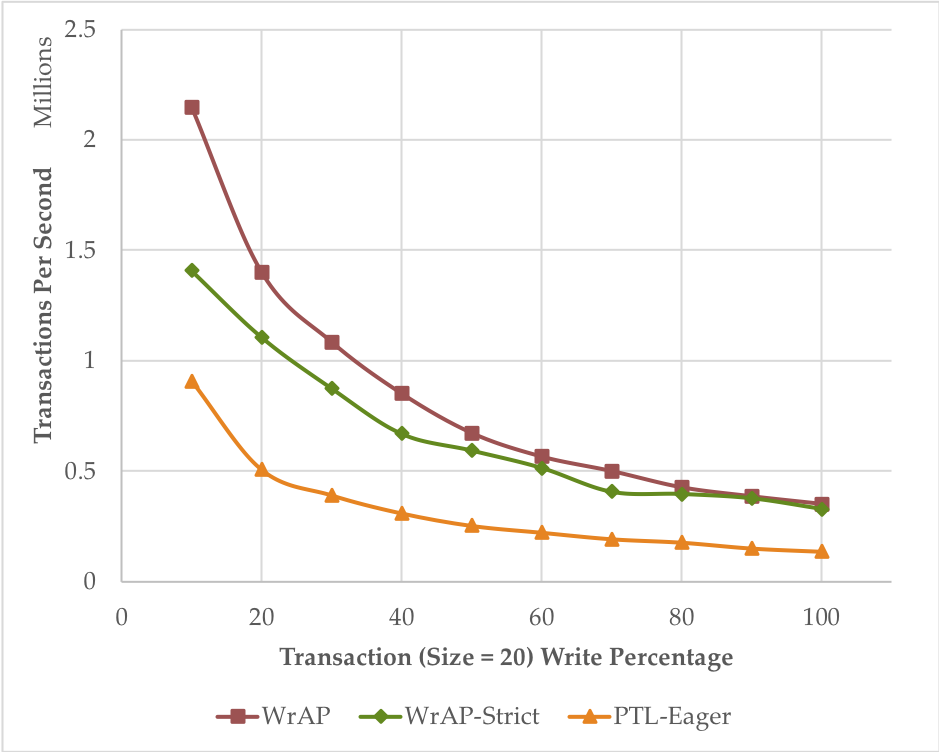}
\caption{Average Txps for Write / Read Percentage of Atomic Hash Table Updates with 6 Concurrent Threads}
\label{fig:s36c}
\end{figure}

The transaction write set was then varied from 2 to 30 elements with 6 concurrent threads.
The average throughput was recorded and is shown in Figure ~\ref{fig:s36}.
Even with adding strict durability, WrAP performs roughly three times faster than {\small PTL-Eager}.

A transaction size of ten elements was then varied with a write to read ratio with 6 concurrent threads.
The average throughput was recorded and is shown in Figure ~\ref{fig:s36c}.
Unlike transactional memory approaches, our approach does not require instrumenting read accesses and can therefore execute reads at cache speeds.

\subsection {Red-Black Tree}

\begin{figure}[th]
\centering
\includegraphics[width=\plotwidth,height=\plotheight]{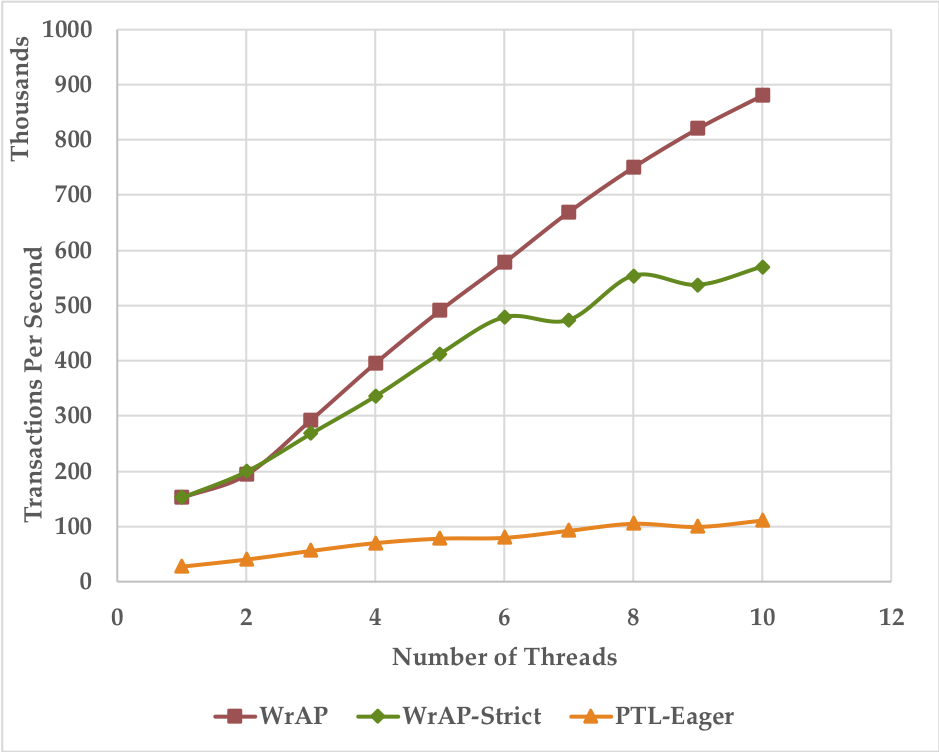}
\caption{Millions of Transactions per Second for Atomic Red-Black Tree Element Inserts versus Number of Concurrent Threads}
\label{fig:s38}
\end{figure}

We use the transactional Red-Black tree from {\small STAMP} ~\cite{stamp} initialized with 1 million elements.
We then perform insert operations on the Red-Black tree and record average transaction times and throughput over 200k additional inserts.
Each transaction inserts an additional element into the Red-Black tree.
Inserting an element into a Red-Black tree first requires finding the insertion point which can take many read operations and can trigger many writes through a rebalance.
In our experiments, we averaged 63 reads and 11 writes per transactional insert of one element into the Red-Black tree.

We record the maximum throughput of inserts into the Red-Black tree per second for a varying number of threads in Figure ~\ref{fig:s38}.
As can be seen in the Figure, WrAP has almost 9x higher throughput over {\small PTL-Eager}, and with strict durability almost 6x faster.
Our method can perform reads at the speed of the hardware, while {\small PTL-Eager} requires instrumenting reads through software to track dependencies on other concurrent transactions.

\subsection {\PMC Analysis}
\label{sec:simulation}

We investigated the required length of our {\small FIFO} in the Volatile Delay Buffer and performance with respect to \PMemory write times using an approach similar to ~\cite{zhao13}.
In the absence of readily available memory controllers, we modified the McSimA+ simulator~\cite{ahn2013}.
McSimA+ is a PIN~\cite{pin05} based simulator that decouples execution from simulation and tightly models out-of-order processor micro-architecture at the cycle level.
We extended the simulator to support the notifications for opening and closing WrAPs along with extended support for memory reads and writes.
We added support for {\small DRAMS}im2 ~\cite{dramsim2}, a cycle-accurate memory system and {\small DRAM} memory controller model library.
Write-combining and store buffers were then added with multiple configuration options to allow fine tuning to match the system to be modeled.

\begin{figure}[t]
\centering
\includegraphics[width=\plotwidth,height=\plotheight]{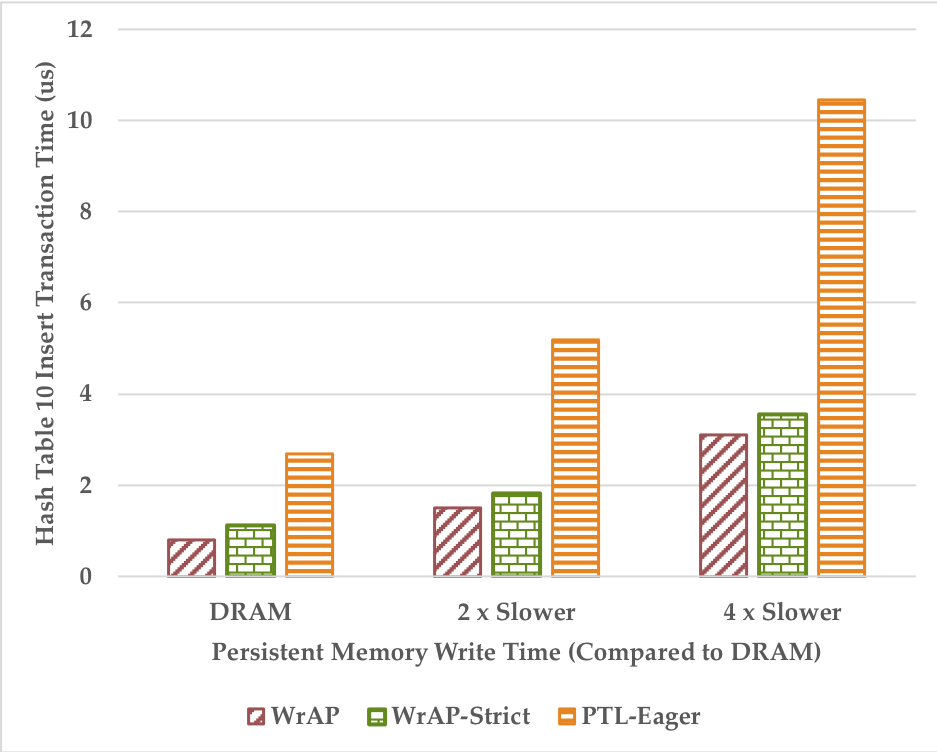}
\caption{Average Atomic Hash Table 10 Element Update with Various \PMemory Write Times with 8 Concurrent Threads}
\label{fig:h1}
\end{figure}

\begin{figure}[t]
\centering
\includegraphics[width=\plotwidth,height=\plotheight]{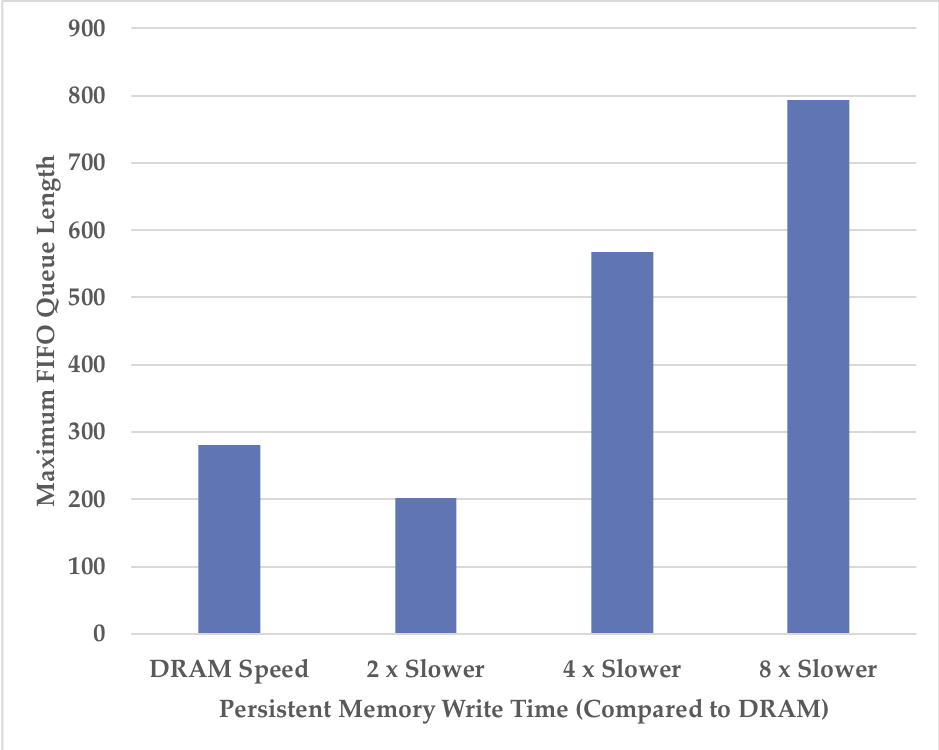}
\caption{Maximum {\small FIFO} Queue Length for Atomic Hash Table 10 Element Update with 4 Concurrent Threads}
\label{fig:h3}
\end{figure}

\begin{figure}[t]
\centering
\includegraphics[width=\plotwidth,height=\plotheight]{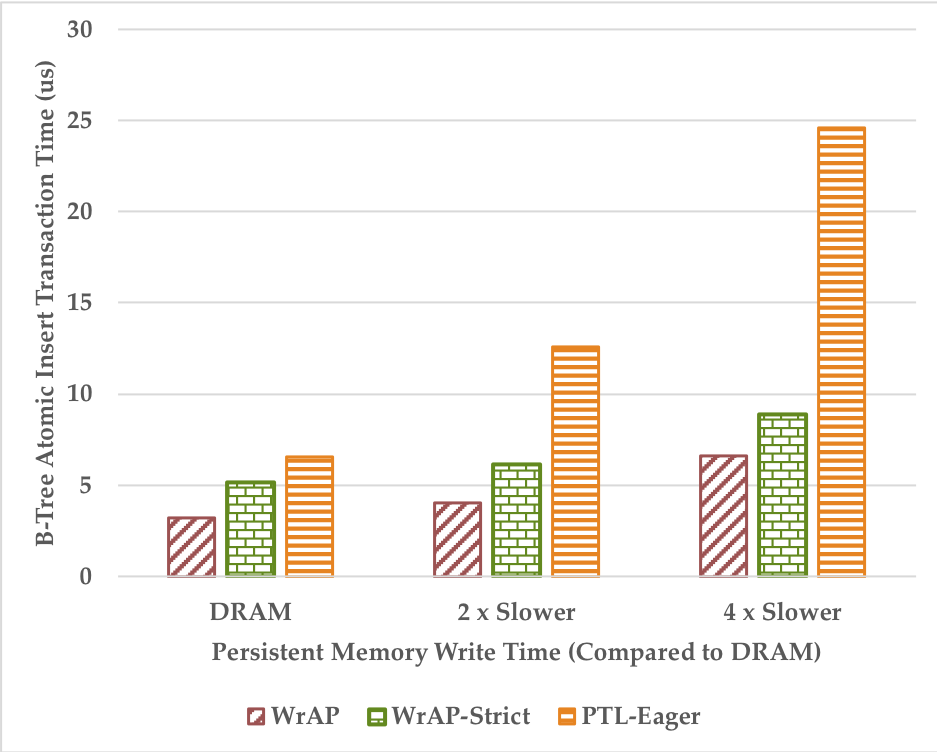}
\caption{Average B-Tree Atomic Element Insert with Various \PMemory Write Times with 8 Concurrent Threads}
\label{fig:h11}
\end{figure}

\begin{figure}[t]
\centering
\includegraphics[width=\plotwidth,height=\plotheight]{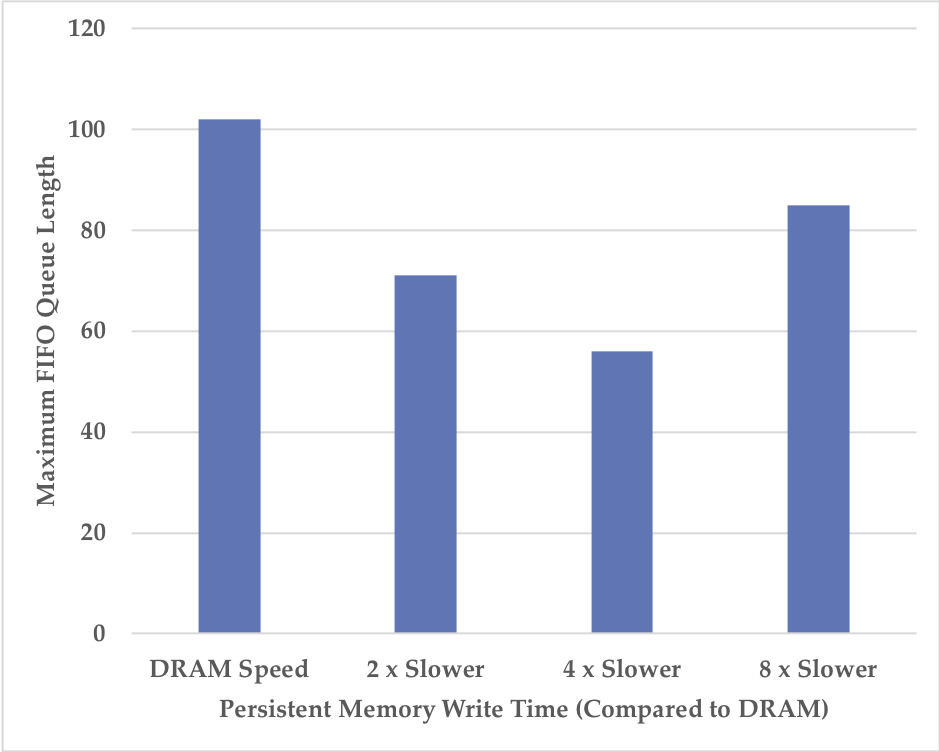}
\caption{Maximum {\small FIFO} Queue Length for B-Tree Atomic Element Insert with 8 Concurrent Threads}
\label{fig:h13}
\end{figure}

To stress the \PMCN, we executed an atomic hash table update without any thread contention by having each thread update elements on a separate portion of the table.
In the simulation, we fill the cache with dirty cache lines so that each write by a thread in a transaction generates write-backs to main \PMemoryN.
For 8 threads, we recorded the average atomic hash table update time for 10 elements in each transaction.
We then vary the \PMemory write time as a multiple of {\small DRAM} write time.
As shown in Figure ~\ref{fig:h1}, WrAP is less affected by increasing write times when compared to {\small PTL-Eager}.

Additionally, we record the maximum {\small FIFO} buffer size for various \PMemory write times and 4 concurrent threads, shown in Figure ~\ref{fig:h3}.
Initially, the buffer size decreases for an increasing \PM write time, due to slower transaction throughput and less cache evictions into the buffer.
As the write time increases, the buffer length increases, but is still less than 1k cache lines or 64{\small KB}.

We performed a similar analysis using a B-Tree, where each thread atomically inserts elements on its own copy of a B-Tree.
Each insert into the tree required, on average, over 5 times as many reads as writes.
As shown in Figure ~\ref{fig:h11}, our method is less affected by increasing \PM write times, due to {\small PTL-Eager} instrumenting the large portion of the read operations.
In this experiment, we use eight concurrent threads each atomically inserting elements into an initialized B-Tree of 128 elements.

As more reads than writes are generated for each atomic insert transaction, the {\small FIFO} buffer length remains small.
We also examined the {\small FIFO} buffer length in the {\small VDB} with 8 concurrent threads.
Figure ~\ref{fig:h13} shows the length was less than about 100 elements for each write speed due to the large proportion of reads.
\vspace{9pt}

\section{Related Work} \label{sec:previous}

{\bf Related Persistence Work:} 
Analysis of consistency models for persistent memory was considered in~\cite{pelley2014}.
Changes to the front-end cache for ordering cache evictions were proposed in~\cite{condit09,venkat11,zhao13,atom}. 
BPFS~\cite{condit09} proposed {\it epoch barriers} to control eviction order, while~\cite{venkat11} proposed  a {\it flush} software primitive to control of update order.
Snapshotting the entire micro architectural state at the point of a failure is proposed in ~\cite{narayanan12}.
A non-volatile victim cache to provide transactional buffering was proposed in~\cite{zhao13}, with the added property of not requiring logging, but requires changes to the front-end cache controller to track pre- and post- transactional states for cache lines in both volatile and persistent caches, atomically moving them to durable state on transaction commits.

Memory controller support for transaction atomicity in \PMemory have been proposed in ~\cite{qureshiISCA2009,zhou2010,zhao2014, cf13, cal16, doshi16, spaa17}.
Adding a small DRAM buffer in front of \PMemory to improve latency and to coalesce writes was proposed in~\cite{qureshiISCA2009}.
The use of a volatile victim cache to prevent uncontrolled cache evictions from reaching \PM was described in
~\cite{cf13,cal16,doshi16}, 
but requires software locking for concurrency control.
FIRM~\cite{zhao2014} describes techniques to differentiate persistent and non-persistent memory traffic, and presents scheduling algorithms to maximize system throughput and fairness.
Low-level memory scheduling to improve efficiency of persistent memory access was studied in~\cite{zhou2010}.
Except for~\cite{cf13,cal16,doshi16}, none of these works deal with the issues of atomicity or durability of write sequences.
Our approach effectively uses {\small HTM} for concurrency control and does not require changes to the font-end cache controller or use logs for replaying transactions to \PMN.

{\bf Related Concurrency Work:}
Existing non-HTM solutions~\cite{volos11,atlas14,rewind} tightly couple concurrency control with durable writes of either write-ahead logs or data updates into \PMemory to maintain persistence consistency.
Software that employs these approaches generally means they must extend the duration for which they remain in critical sections, leading to longer times to hold locks, which reduces concurrency and expands transactional duration.
Other work~\cite{msst15, dudetm} decouples concurrency control so that post transactional values may flow through cache hierarchy and reach \PM asynchronously; however, the write ahead log for an updating transaction has to get committed into \PM synchronously before the transaction can close so that the integrity of the foreground value flow is preserved across machine restarts.
Another hardware-assisted mechanism proposes hardware changes to allow a dual-scheme checkpointing that writes previous check-pointed values in the background while collecting current transaction writes ~\cite{thynvm}.

Recent work~\cite{avni1,avni2,dudetm} aims to exploit processor-supported {\small HTM} mechanisms for concurrency control instead of traditional locking or {\small STM}-based approaches. 
However, all of these solutions require making significant changes to the existing {\small HTM} semantics and implementations.
For instance, {\small PHTM}~\cite{avni1} and {\small PHyTM}~\cite{avni2}, propose a new instruction called {\bf TransparentFlush} which can be used to flush a cache line from {\it within a transaction} to persistent memory without causing any transaction to abort.
They also propose a change to the {\bf xend} instruction that ends an atomic {\small HTM} region, so that it atomically updates a bit in persistent memory as part of its execution.
Similarly, for {\small DUDETM}~\cite{dudetm} to use {\small HTM}, it requires that designated memory variables {\it within a transaction} be allowed to be updated globally and concurrently without causing an abort.
Other work~\cite{ismm17} utilizes {\small HTM} for concurrency control, but requires aliasing all read and write accesses while concurrently maintaining log ordering and and replaying logs for retirement.

\section{Summary}

In this paper we presented an approach that unifies {\small HTM} and \PM to create durable, concurrent transactions.
Our approach works with existing {\small HTM} and cache coherency mechanisms, and does not require changes to existing processor caches or store instructions, avoids synchronous cache-line write-backs on completions, and only utilizes logs for recovery.
The solution correctly orders {\small HTM} transactions and atomically commits them to \PMemory by the use of a novel software protocol combined with a back-end \PMCN.

Our approach, evaluated using both micro-benchmarks and the {\small STAMP} suite compares well with standard (volatile) {\small HTM} transactions.
In comparison with persistent transactional locking, our approach performs 3x faster on standard benchmarks and almost 9x faster on a Red-Black Tree data structure.

\bibliographystyle{IEEEtran}
\bibliography{paper}

\begin{thebibliography}{10}
\providecommand{\url}[1]{#1}
\csname url@samestyle\endcsname
\providecommand{\newblock}{\relax}
\providecommand{\bibinfo}[2]{#2}
\providecommand{\BIBentrySTDinterwordspacing}{\spaceskip=0pt\relax}
\providecommand{\BIBentryALTinterwordstretchfactor}{4}
\providecommand{\BIBentryALTinterwordspacing}{\spaceskip=\fontdimen2\font plus
\BIBentryALTinterwordstretchfactor\fontdimen3\font minus
  \fontdimen4\font\relax}
\providecommand{\BIBforeignlanguage}[2]{{%
\expandafter\ifx\csname l@#1\endcsname\relax
\typeout{** WARNING: IEEEtran.bst: No hyphenation pattern has been}%
\typeout{** loaded for the language `#1'. Using the pattern for}%
\typeout{** the default language instead.}%
\else
\language=\csname l@#1\endcsname
\fi
#2}}
\providecommand{\BIBdecl}{\relax}
\BIBdecl

\bibitem{saphana}
F.~F\"{a}rber, S.~K. Cha, J.~Primsch, C.~Bornh\"{o}vd, S.~Sigg, and W.~Lehner,
  ``{SAP} {HANA} {D}atabase: {D}ata management for modern business
  applications,'' \emph{SIGMOD Rec.}, vol.~40, no.~4, pp. 45--51, Jan. 2012.

\bibitem{blu13vldb}
V.~Raman, G.~Attaluri, R.~Barber, N.~Chainani, D.~Kalmuk, V.~KulandaiSamy,
  J.~Leenstra, S.~Lightstone, S.~Liu, G.~M. Lohman \emph{et~al.}, ``Db2 with
  blu acceleration: So much more than just a column store,'' \emph{Proceedings
  of the VLDB Endowment}, vol.~6, no.~11, pp. 1080--1091, 2013.

\bibitem{scispark}
R.~Palamuttam, R.~M. Mogrovejo, C.~Mattmann, B.~Wilson, K.~Whitehall, R.~Verma,
  L.~McGibbney, and P.~Ramirez, ``Scispark: Applying in-memory distributed
  computing to weather event detection and tracking,'' in \emph{Big Data (Big
  Data), 2015 IEEE International Conference on}.\hskip 1em plus 0.5em minus
  0.4em\relax IEEE, 2015, pp. 2020--2026.

\bibitem{gridgain}
G.~Team, ``Gridgain: In-memory computing platform,'' 2007.

\bibitem{sparkML}
X.~Meng, J.~Bradley, B.~Yavuz, E.~Sparks, S.~Venkataraman, D.~Liu, J.~Freeman,
  D.~Tsai, M.~Amde, S.~Owen \emph{et~al.}, ``{MLlib}: Machine learning in
  apache spark,'' \emph{Journal of Machine Learning Research}, vol.~17, no.~34,
  pp. 1--7, 2016.

\bibitem{intel3dxp}
\BIBentryALTinterwordspacing
I.~Corporation. (2015, July) Intel and {M}icron {P}roduce {B}reakthrough
  {M}emory {T}echnology. [Online]. Available:
  \url{https://newsroom.intel.com/news-releases/intel-and-micron-produce-breakthrough-memory-technology/}
\BIBentrySTDinterwordspacing

\bibitem{htm}
M.~Herlihy and J.~E.~B. Moss, \emph{Transactional {M}emory: {A}rchitectural
  support for lock-free data structures}.\hskip 1em plus 0.5em minus
  0.4em\relax ACM, 1993, vol. 21,2.

\bibitem{sle}
R.~Rajwar and J.~R. Goodman, ``Speculative lock elision: Enabling highly
  concurrent multithreaded execution,'' in \emph{Proceedings of the 34th annual
  ACM/IEEE international symposium on Microarchitecture}.\hskip 1em plus 0.5em
  minus 0.4em\relax IEEE Computer Society, 2001, pp. 294--305.

\bibitem{volos11}
H.~Volos, A.~J. Tack, and M.~Swift, ``Mnemosyne: Lightweight persistent
  memory,'' in \emph{Proceedings of 16th International Conference on
  Architectural Support for Programming Languages and Operating Systems}.\hskip
  1em plus 0.5em minus 0.4em\relax ACM Press, 2011, pp. 91--104.

\bibitem{msst15}
E.~Giles, K.~Doshi, and P.~Varman, ``{SoftWrAP}: {A} lightweight framework for
  transactional support of storage class memory,'' in \emph{Mass Storage
  Systems and Technologies (MSST), 2015 31st Symposium on}, May 2015, pp.
  1--14.

\bibitem{atlas14}
\BIBentryALTinterwordspacing
D.~R. Chakrabarti, H.-J. Boehm, and K.~Bhandari, ``Atlas: Leveraging locks for
  non-volatile memory consistency,'' in \emph{Proceedings of the 2014 ACM
  International Conference on Object Oriented Programming Systems Languages \&
  Applications}, ser. OOPSLA '14.\hskip 1em plus 0.5em minus 0.4em\relax New
  York, NY, USA: ACM, 2014, pp. 433--452. [Online]. Available:
  \url{http://doi.acm.org/10.1145/2660193.2660224}
\BIBentrySTDinterwordspacing

\bibitem{rewind}
A.~Chatzistergiou, M.~Cintra, and S.~D. Viglas, ``Rewind: Recovery write-ahead
  system for in-memory non-volatile data-structures,'' \emph{Proceedings of the
  VLDB Endowment}, vol.~8, no.~5, pp. 497--508, 2015.

\bibitem{dudetm}
\BIBentryALTinterwordspacing
M.~Liu, M.~Zhang, K.~Chen, X.~Qian, Y.~Wu, W.~Zheng, and J.~Ren, ``{DudeTM:}
  {B}uilding {D}urable {T}ransactions with {D}ecoupling for {P}ersistent
  {M}emory,'' in \emph{Proceedings of the Twenty-Second International
  Conference on Architectural Support for Programming Languages and Operating
  Systems}, ser. ASPLOS '17.\hskip 1em plus 0.5em minus 0.4em\relax New York,
  NY, USA: ACM, 2017, pp. 329--343. [Online]. Available:
  \url{http://doi.acm.org/10.1145/3037697.3037714}
\BIBentrySTDinterwordspacing

\bibitem{zhao13}
\BIBentryALTinterwordspacing
J.~Zhao, S.~Li, D.~H. Yoon, Y.~Xie, and N.~P. Jouppi, ``Kiln: Closing the
  performance gap between systems with and without persistence support,'' in
  \emph{Proceedings of the 46th Annual IEEE/ACM International Symposium on
  Microarchitecture}, ser. MICRO-46.\hskip 1em plus 0.5em minus 0.4em\relax New
  York, NY, USA: ACM, 2013, pp. 421--432. [Online]. Available:
  \url{http://doi.acm.org/10.1145/2540708.2540744}
\BIBentrySTDinterwordspacing

\bibitem{atom}
A.~Joshi, V.~Nagarajan, S.~Viglas, and M.~Cintra, ``{ATOM}: Atomic durability
  in non-volatile memory through hardware logging,'' in \emph{2017 IEEE
  International Symposium on High Performance Computer Architecture (HPCA)},
  2017.

\bibitem{spms}
\BIBentryALTinterwordspacing
S.~Li, P.~Wang, N.~Xiao, G.~Sun, and F.~Liu, ``{SPMS}: Strand based persistent
  memory system,'' in \emph{Proceedings of the Conference on Design, Automation
  \& Test in Europe}, ser. DATE '17.\hskip 1em plus 0.5em minus 0.4em\relax
  3001 Leuven, Belgium, Belgium: European Design and Automation Association,
  2017, pp. 622--625. [Online]. Available:
  \url{http://dl.acm.org/citation.cfm?id=3130379.3130529}
\BIBentrySTDinterwordspacing

\bibitem{cf13}
E.~Giles, K.~Doshi, and P.~Varman, ``Bridging the programming gap between
  persistent and volatile memory using {WrAP},'' in \emph{Proceedings of the
  ACM International Conference on Computing Frontiers}.\hskip 1em plus 0.5em
  minus 0.4em\relax ACM, 2013, p.~30.

\bibitem{cal16}
L.~Pu, K.~Doshi, E.~Giles, and P.~Varman, ``Non-{I}ntrusive {P}ersistence with
  a {B}ackend {NVM} {C}ontroller,'' \emph{IEEE Computer Architecture Letters},
  vol.~15, no.~1, pp. 29--32, Jan 2016.

\bibitem{doshi16}
K.~Doshi, E.~Giles, and P.~Varman, ``Atomic {P}ersistence for {SCM} with a
  {N}on-intrusive {B}ackend {C}ontroller,'' in \emph{The 22nd International
  Symposium on High-Performance Computer Architecture (HPCA)}.\hskip 1em plus
  0.5em minus 0.4em\relax IEEE, March 2016.

\bibitem{ptm}
Z.~Wang, H.~Yi, R.~Liu, M.~Dong, and H.~Chen, ``Persistent transactional
  memory,'' \emph{IEEE Computer Architecture Letters}, vol.~14, no.~1, pp.
  58--61, Jan 2015.

\bibitem{avni1}
H.~Avni, E.~Levy, and A.~Mendelson, ``Hardware transactions in nonvolatile
  memory,'' in \emph{Proceedings of the 29th International Symposium on
  Distributed Computing - Volume 9363}, ser. DISC 2015.\hskip 1em plus 0.5em
  minus 0.4em\relax New York, NY, USA: Springer-Verlag New York, Inc., 2015,
  pp. 617--630.

\bibitem{avni2}
H.~Avni and T.~Brown, ``{PHyTM:} {P}ersistent hybrid transactional memory,''
  \emph{Proceedings of the VLDB Endowment}, vol.~10, no.~4, pp. 409--420, 2016.

\bibitem{spaa17}
\BIBentryALTinterwordspacing
E.~Giles, K.~Doshi, and P.~Varman, ``Brief announcement: Hardware transactional
  storage class memory,'' in \emph{Proceedings of the 29th ACM Symposium on
  Parallelism in Algorithms and Architectures}, ser. SPAA '17.\hskip 1em plus
  0.5em minus 0.4em\relax New York, NY, USA: ACM, 2017, pp. 375--378. [Online].
  Available: \url{http://doi.acm.org/10.1145/3087556.3087589}
\BIBentrySTDinterwordspacing

\bibitem{spear13}
W.~Ruan, Y.~Liu, and M.~Spear, ``Boosting timestamp-based transactional memory
  by exploiting hardware cycle counters,'' \emph{ACM Transactions on
  Architecture and Code Optimization (TACO)}, vol.~10, no.~4, p.~40, 2013.

\bibitem{tsxprof}
Y.~Liu, J.~Gottschlich, G.~Pokam, and M.~Spear, ``Tsxprof: Profiling hardware
  transactions,'' in \emph{Parallel Architecture and Compilation ($PACT$), 2015
  International Conference on}.\hskip 1em plus 0.5em minus 0.4em\relax IEEE,
  2015, pp. 75--86.

\bibitem{ismm17}
\BIBentryALTinterwordspacing
E.~Giles, K.~Doshi, and P.~Varman, ``Continuous {C}heckpointing of {HTM}
  {T}ransactions in {NVM},'' in \emph{Proceedings of the 2017 ACM SIGPLAN
  International Symposium on Memory Management}, ser. {ISMM} 2017.\hskip 1em
  plus 0.5em minus 0.4em\relax New York, NY, USA: ACM, 2017, pp. 70--81.
  [Online]. Available: \url{http://doi.acm.org/10.1145/3092255.3092270}
\BIBentrySTDinterwordspacing

\bibitem{IntelTSX}
{Intel Corporation}, ``{I}ntel {T}ransactional {S}ynchronization
  {E}xtensions,'' in \emph{{I}ntel {A}rchitecture {I}nstruction {S}et
  {E}xtensions {P}rogramming {R}eference}, February 2012, ch.~8,
  http://software.intel.com/.

\bibitem{ssca}
D.~A. Bader and K.~Madduri, ``Design and implementation of the {HPCS} graph
  analysis benchmark on symmetric multiprocessors,'' in \emph{International
  Conference on High-Performance Computing}.\hskip 1em plus 0.5em minus
  0.4em\relax Springer, 2005, pp. 465--476.

\bibitem{stamp}
C.~C. Minh, J.~Chung, C.~Kozyrakis, and K.~Olukotun, ``{STAMP}: Stanford
  transactional applications for multi-processing,'' in \emph{Workload
  Characterization, 2008. IISWC 2008. IEEE International Symposium on}.\hskip
  1em plus 0.5em minus 0.4em\relax IEEE, 2008, pp. 35--46.

\bibitem{TL}
D.~Dice and N.~Shavit, ``Understanding tradeoffs in software transactional
  memory,'' in \emph{Code Generation and Optimization, 2007. CGO'07.
  International Symposium on}.\hskip 1em plus 0.5em minus 0.4em\relax IEEE,
  2007, pp. 21--33.

\bibitem{TL2}
D.~Dice, O.~Shalev, and N.~Shavit, ``Transactional {L}ocking {II},'' in
  \emph{Distributed Computing}.\hskip 1em plus 0.5em minus 0.4em\relax
  Springer, 2006, pp. 194--208.

\bibitem{tl2x86}
\BIBentryALTinterwordspacing
C.~C. Minh, ``{TL2-x86}, a port of tl2 to x86 architecture,'' in \emph{On
  GitHub, ccaominh, tl2-x86}.\hskip 1em plus 0.5em minus 0.4em\relax Stanford,
  2015. [Online]. Available: \url{https://github.com/ccaominh/tl2-x86}
\BIBentrySTDinterwordspacing

\bibitem{ahn2013}
J.~H. Ahn, S.~Li, O.~Seongil, and N.~P. Jouppi, ``Mcsima+: A manycore simulator
  with application-level+ simulation and detailed microarchitecture modeling,''
  in \emph{Performance Analysis of Systems and Software (ISPASS), 2013 IEEE
  International Symposium on}.\hskip 1em plus 0.5em minus 0.4em\relax IEEE,
  2013, pp. 74--85.

\bibitem{pin05}
C.-K. Luk, R.~Cohn, R.~Muth, H.~Patil, A.~Klauser, G.~Lowney, S.~Wallace, V.~J.
  Reddi, and K.~Hazelwood, ``Pin: building customized program analysis tools
  with dynamic instrumentation,'' in \emph{ACM Sigplan Notices}, vol.~40,
  no.~6.\hskip 1em plus 0.5em minus 0.4em\relax ACM, 2005, pp. 190--200.

\bibitem{dramsim2}
P.~Rosenfeld, E.~Cooper-Balis, and B.~Jacob, ``Dramsim2: A cycle accurate
  memory system simulator,'' \emph{Computer Architecture Letters}, vol.~10,
  no.~1, pp. 16--19, 2011.

\bibitem{pelley2014}
S.~Pelley, P.~M. Chen, and T.~F. Wenisch, ``Memory persistency,'' in
  \emph{Computer Architecture (ISCA), 2014 ACM/IEEE 41st International
  Symposium on}.\hskip 1em plus 0.5em minus 0.4em\relax IEEE, 2014, pp.
  265--276.

\bibitem{condit09}
\BIBentryALTinterwordspacing
J.~Condit, E.~B. Nightingale, C.~Frost, E.~Ipek, B.~Lee, D.~Burger, and
  D.~Coetzee, ``Better {I/O} through byte-addressable, persistent memory,'' in
  \emph{Proceedings of the ACM SIGOPS 22Nd Symposium on Operating Systems
  Principles}, ser. SOSP '09.\hskip 1em plus 0.5em minus 0.4em\relax New York,
  NY, USA: ACM, 2009, pp. 133--146. [Online]. Available:
  \url{http://doi.acm.org/10.1145/1629575.1629589}
\BIBentrySTDinterwordspacing

\bibitem{venkat11}
S.~Venkatraman, N.~Tolia, P.~Ranganathan, and R.~H. Campbell, ``Consistent and
  durable data structures for non-volatile byte addressable memory,'' in
  \emph{Proceedings of 9th Usenix Conference on File and Storage
  Technologies}.\hskip 1em plus 0.5em minus 0.4em\relax ACM Press, 2011, pp.
  61--76.

\bibitem{narayanan12}
D.~Narayanan and O.~Hodson, ``Whole-system persistence,'' in \emph{Proceedings
  of 17th International Conference on Architectural Support for Programming
  Languages and Operating Systems}.\hskip 1em plus 0.5em minus 0.4em\relax ACM
  Press, 2012, pp. 401--410.

\bibitem{qureshiISCA2009}
\BIBentryALTinterwordspacing
M.~K. Qureshi, V.~Srinivasan, and J.~A. Rivers, ``Scalable high performance
  main memory system using phase-change memory technology,'' in
  \emph{Proceedings of the 36th Annual International Symposium on Computer
  Architecture}, ser. ISCA '09.\hskip 1em plus 0.5em minus 0.4em\relax New
  York, NY, USA: ACM, 2009, pp. 24--33. [Online]. Available:
  \url{http://doi.acm.org/10.1145/1555754.1555760}
\BIBentrySTDinterwordspacing

\bibitem{zhou2010}
P.~Zhou, Y.~Du, Y.~Zhang, and J.~Yang, ``Fine-grained {Q}o{S} scheduling for
  {PCM}-based main memory systems,'' in \emph{Parallel \& Distributed
  Processing (IPDPS), 2010 IEEE International Symposium on}.\hskip 1em plus
  0.5em minus 0.4em\relax IEEE, 2010, pp. 1--12.

\bibitem{zhao2014}
J.~Zhao, O.~Mutlu, and Y.~Xie, ``Firm: Fair and high-performance memory control
  for persistent memory systems,'' in \emph{Microarchitecture (MICRO), 2014
  47th Annual IEEE/ACM International Symposium on}.\hskip 1em plus 0.5em minus
  0.4em\relax IEEE, 2014, pp. 153--165.

\bibitem{thynvm}
J.~Ren, J.~Zhao, S.~Khan, J.~Choi, Y.~Wu, and O.~Mutlu, ``{ThyNVM}: Enabling
  software-transparent crash consistency in persistent memory systems,'' in
  \emph{Proceedings of the 48th International Symposium on
  Microarchitecture}.\hskip 1em plus 0.5em minus 0.4em\relax ACM, 2015, pp.
  672--685.

\end{thebibliography}

\end{document}